\documentclass{aastex631}

\usepackage{amsmath}
\usepackage{subfigure}
\usepackage{textcomp}
\usepackage{hyperref}
\usepackage[misc]{ifsym} 
\shortauthors{Tao et al.}
\graphicspath{{./}{figures/}}

\begin{document}

\title{Sensitive Multi-beam Targeted SETI Observations towards 33 Exoplanet Systems with FAST}

\author[0000-0002-4683-5500]{Zhen-Zhao Tao}
\altaffiliation{These authors contributed equally to this work.}
\affiliation{Institute for Frontiers in Astronomy and Astrophysics, Beijing Normal University, Beijing 102206, China}
\affiliation{Department of Astronomy, Beijing Normal University, Beijing 100875, China; \url{tjzhang@bnu.edu.cn}}
\affiliation{Institute for Astronomical Science, Dezhou University, Dezhou 253023, China}

\author[0000-0002-5485-1877]{Hai-Chen Zhao}
\altaffiliation{These authors contributed equally to this work.}
\affiliation{Institute for Frontiers in Astronomy and Astrophysics, Beijing Normal University, Beijing 102206, China}
\affiliation{Department of Astronomy, Beijing Normal University, Beijing 100875, China; \url{tjzhang@bnu.edu.cn}}


\author[0000-0002-3363-9965]{Tong-Jie Zhang\href{mailto:tjzhang@bnu.edu.cn}{\textrm{\Letter}}}
\affiliation{Institute for Frontiers in Astronomy and Astrophysics, Beijing Normal University, Beijing 102206, China}
\affiliation{Department of Astronomy, Beijing Normal University, Beijing 100875, China; \url{tjzhang@bnu.edu.cn}}
\affiliation{Institute for Astronomical Science, Dezhou University, Dezhou 253023, China}

\author[0000-0002-8604-106X]{Vishal Gajjar\href{mailto:vishalg@berkeley.edu}{\textrm{\Letter}}}
\affiliation{Breakthrough Listen, University of California Berkeley, Berkeley, CA 94720, USA; \url{vishalg@berkeley.edu}}


\author{Yan Zhu}
\affiliation{National Astronomical Observatories, Chinese Academy of Sciences, Beijing 100012, China; \url{dili@nao.cas.cn}}

\author[0000-0003-4415-2148]{You-Ling Yue}
\affiliation{National Astronomical Observatories, Chinese Academy of Sciences, Beijing 100012, China; \url{dili@nao.cas.cn}}

\author{Hai-Yan Zhang}
\affiliation{National Astronomical Observatories, Chinese Academy of Sciences, Beijing 100012, China; \url{dili@nao.cas.cn}}

\author{Wen-Fei Liu}
\affiliation{College of Physics and Electronic Engineering, Qilu Normal University, Jinan 250200, China}

\author{Shi-Yu Li}
\affiliation{Beijing Planetarium, Beijing Academy of Science and Technology, Beijing 100044, China}

\author{Jian-Chen Zhang}
\affiliation{Institute for Astronomical Science, Dezhou University, Dezhou 253023, China}

\author{Cong Liu}
\affiliation{Institute for Astronomical Science, Dezhou University, Dezhou 253023, China}

\author{Hong-Feng Wang}
\affiliation{Institute for Astronomical Science, Dezhou University, Dezhou 253023, China}

\author[0000-0002-9260-1641]{Ran Duan}
\affiliation{National Astronomical Observatories, Chinese Academy of Sciences, Beijing 100012, China; \url{dili@nao.cas.cn}}

\author[0000-0003-0597-0957]{Lei Qian}
\affiliation{National Astronomical Observatories, Chinese Academy of Sciences, Beijing 100012, China; \url{dili@nao.cas.cn}}

\author{Cheng-Jin Jin}
\affiliation{National Astronomical Observatories, Chinese Academy of Sciences, Beijing 100012, China; \url{dili@nao.cas.cn}}

\author[0000-0003-3010-7661]{Di Li\href{mailto:dili@nao.cas.cn}{\textrm{\Letter}}}
\affiliation{National Astronomical Observatories, Chinese Academy of Sciences, Beijing 100012, China; \url{dili@nao.cas.cn}}

\author{Andrew Siemion}
\affiliation{Breakthrough Listen, University of California Berkeley, Berkeley, CA 94720, USA; \url{vishalg@berkeley.edu}}

\author{Peng Jiang}
\affiliation{National Astronomical Observatories, Chinese Academy of Sciences, Beijing 100012, China; \url{dili@nao.cas.cn}}

\author{Dan Werthimer\href{mailto:danw@ssl.berkeley.edu}{\textrm{\Letter}}}
\affiliation{Breakthrough Listen, University of California Berkeley, Berkeley, CA 94720, USA; \url{vishalg@berkeley.edu}}
\affiliation{Space Sciences Laboratory, University of California Berkeley, Berkeley, CA 94720, USA; \url{danw@ssl.berkeley.edu}}

\author{Jeff Cobb}
\affiliation{Breakthrough Listen, University of California Berkeley, Berkeley, CA 94720, USA; \url{vishalg@berkeley.edu}}
\affiliation{Space Sciences Laboratory, University of California Berkeley, Berkeley, CA 94720, USA; \url{danw@ssl.berkeley.edu}}

\author{Eric Korpela}
\affiliation{Space Sciences Laboratory, University of California Berkeley, Berkeley, CA 94720, USA; \url{danw@ssl.berkeley.edu}}

\author{David P. Anderson}
\affiliation{Space Sciences Laboratory, University of California Berkeley, Berkeley, CA 94720, USA; \url{danw@ssl.berkeley.edu}}



\begin{abstract}

As a major approach to looking for life beyond the Earth, the search for extraterrestrial intelligence (SETI) is committed to searching for technosignatures such as engineered radio signals that are indicative of technologically capable life. In this paper, we report a targeted SETI campaign employing an observation strategy named multi-beam coincidence matching (MBCM) at the Five-hundred-meter Aperture Spherical radio Telescope (FAST) towards 33 known exoplanet systems, searching for ETI narrow-band drifting signals across $1.05-1.45$ GHz in two orthogonal linear polarization directions separately. A signal at 1140.604 MHz detected from the observation towards Kepler-438 originally piqued our interest because its features are roughly consistent with assumed ETI technosignatures. However, evidences such as its polarization characteristics are able to eliminae the possibility of an extraterrestrial origin. Our observations achieve an unprecedented sensitivity since the minimum equivalent isotropic radiated power (EIRP) we are able to detect reaches $1.48 \times 10^{9} ~\text{W}$.

\end{abstract}

\keywords{\href{http://astrothesaurus.org/uat/74}{Astrobiology (74)}; \href{http://astrothesaurus.org/uat/2127}{Search for extraterrestrial intelligence (2127)}; \href{http://astrothesaurus.org/uat/2128}{Technosignatures (2128)}; \href{http://astrothesaurus.org/uat/498}{Exoplanets (498)}}


\section{Introduction} \label{sec:intro}

    ``Are we alone in the universe?'' This has been one of the most mysterious and profound questions in astronomy since ancient times. To answer this question, there are three primary methods to search for life beyond the Earth: detection in situ of biosignatures (life and byproducts of biological processes) at sites of interest; remote sensing of biosignatures from planetary atmospheres; and the detection of technosignatures (signs of technologically sophisticated civilizations), i.e. SETI \citep{1959Natur.184..844C}. Since the first experiment in the 1960s \citep{1961PhT....14d..40D}, SETI has been mostly conducted in the radio waveband. In the early years, radio SETI observations focussed on several specific frequencies such as the neutral hydrogen 21-cm line, and the bandwidths were narrower than several megahertz \citep{1973Icar...19..329V,1980Icar...42..136T,1986Icar...65..152V,1993ApJ...415..218H}. The available bandwidth for SETI has now expanded to tens of gigahertz \citep{2018PASP..130d4502M}. 
    
    Radio SETI observations can be conducted in two ways: sky surveys and targeted searches \citep{2001ARA&A..39..511T}. The SERENDIP and SETI@home projects are representatives of radio SETI sky surveys, conducted in the 1990s using the Arecibo radio telescope \citep{2001SPIE.4273..104W}. In recent years, targeted SETI observations has become prevalent due to the discoveries of more and more exoplanets \citep{2013ApJ...767...95D,2013PNAS..11019273P,2014PNAS..11112647B}. Since 2017, the Breakthrough Listen initiative has conducted a series of targeted SETI observations using the Green Bank telescope (GBT) and the Parkes radio telescope (Parkes), whose targets have involved nearby stars \citep{2017ApJ...849..104E,2020AJ....159...86P,2020AJ....160...29S}, discovered exoplanets \citep{2021AJ....161..286T,2021NatAs.tmp..203S}, and the Galactic Center \citep{2021AJ....162...33G}. However, though massive efforts have been made, no conclusive evidence of the existence of extraterrestrial intelligence (ETI) has been found so far. But even though the possibility of a discovery is minimal, if we stop trying, the chance is zero.

Narrow-band ($\sim$ Hz) signals are the most common type targeted by radio SETI. Widely used in human electromagnetic communications, narrow-band signals are indicative of technology because they cannot be produced by any natural astrophysical process, and can arise from either intentional transmission or leakage \citep{2001ARA&A..39..511T}. A narrow-band signal transmitted from a distant source drifts in frequency due to the Doppler effect, and the drift rate is given by $\dot{\nu} = \nu_{0}a/c$, where $\nu_{0}$ is the original frequency 
of the signal, $a$ is the relative acceleration between the transmitter and the receiver, and $c$ is the speed of light. An advanced civilization may also attract others by signals that mimic astrophysical ones, such as signals like fast radio bursts (FRBs) \citep{2020FrPhy..1554502Z}, but at present narrow-band signals are the most distinguishable ETI technosignatures for us.

With an enormous illuminated aperture (300 m), a cryogenically-cooled system temperature ($\sim$20 K) for the L-band ($1.05 - 1.45$ GHz) 19-beam receiver, and a large sky region coverage ($-14^{\circ}$ to $+66^{\circ}$ in declination), FAST \citep{2011IJMPD..20..989N,2016RaSc...51.1060L,2019SCPMA..6259502J} is well positioned to conduct highly sensitive and efficient searches for various astronomical objects such as pulsars \citep{2019SCPMA..6259508Q,2021RAA....21..107H}, FRBs \citep{2020ApJ...895L...6Z,2020Natur.587...63L,2021Natur.598..267L}, and ETI radio technosignatures. SETI is one of the five key science goals specified in the original FAST project plan \citep{2006ScChG..49..129N}. In 2019, the first SETI survey of FAST was performed by commensal observations, where two groups of candidate signals were detected \citep{2020ApJ...891..174Z}. Apart from narrow-band signals, FAST is also capable of searching for two types of broad-band technosignatures: artificially-dispersed pulses and artificially-modulated signals \citep{2020RAA....20...78L}, which will be explored in the future.

Identifying ubiquitous human-generated radio frequency interference (RFI) has always been challenging for radio SETI, especially in this era of increasing radio contamination. For most observations aiming at natural objects, narrow-band RFI should be removed directly from the data \citep{2019SCPMA..6259508Q,2020ApJ...895L...6Z,2021RAA....21..107H,2020Natur.587...63L,2021Natur.598..267L}. But this is not applicable to radio SETI, since the ETI narrow-band signals we are searching for may also be removed by this procedure. In recent years, the most commonly used method for targeted SETI observations to identify RFI is the on-off strategy \citep{2013ApJ...767...94S,2017ApJ...849..104E,2019AJ....157..122P,2020AJ....159...86P,2020AJ....160...29S,2021AJ....161..286T,2021AJ....162...33G,2021NatAs.tmp..203S}. The observations of a target (on-source) are interspersed with pointings towards several reference locations (off-sources) which are chosen to be at least several times the full width at half maximum (FWHM) of the telescope away from the target, thus ensuring that signals from the on-source are unlikely to be detected in the off-observations. In contrast, RFI entering the side lobes of the beam is expected to be present in both the on- and the off-observations and can be rejected immediately. So far the on-off strategy has been applied to GBT and Parkes. However, the arrangement of the 19 beams on the FAST L-band receiver makes it possible to distinguish RFI from sky localized ETI signals without doing on-off observations.

Multi-beam observations are typically performed when conducting large-scale sky surveys to achieve high efficiency. Parkes has been searching for FRBs and pulsars using its 13-beam receiver \citep{2018MNRAS.473..116K}, so has FAST with its 19-beam receiver \citep{2018IMMag..19..112L}. However, multi-beam observations are rarely used for targeted searches, especially for SETI. A targeted SETI study using the Allen Telescope Array (ATA) rejected RFI by pointing three synthesized beams to different stars simultaneously \citep{2016AJ....152..181H}. Compared with large-number small-diameter radio telescope arrays, single-dish large-aperture telescopes have unparalleled advantages in sensitivity. Since a similar strategy has never been used on a single-dish multi-beam telescope for targeted SETI, we design such an observation strategy named multi-beam coincidence matching for FAST.

In this paper, we present the methods and results of the first targeted SETI observations conducted by FAST towards 33 exoplanet systems. In Section \ref{sec:Obser}, we discuss the targets observed in this work and the principles of the MBCM strategy. The techniques we use to analyze our data are discussed in Section \ref{sec:DataAnalysis}. The results of signal search and identification are presented in Section \ref{sec:Results}. In Section \ref{sec: Discussion}, we discuss the sensitivity of our observations, the advantages of our observation strategy, and some complements to the current technosignature verification methodology. Finally, the conclusions of this work are presented in Section \ref{sec:con}.

\section{Observations} \label{sec:Obser}

\subsection{Targets} \label{sec:select}

\begin{figure}[htbp]
\centering
\subfigure[]{
\includegraphics[width=0.48\linewidth]{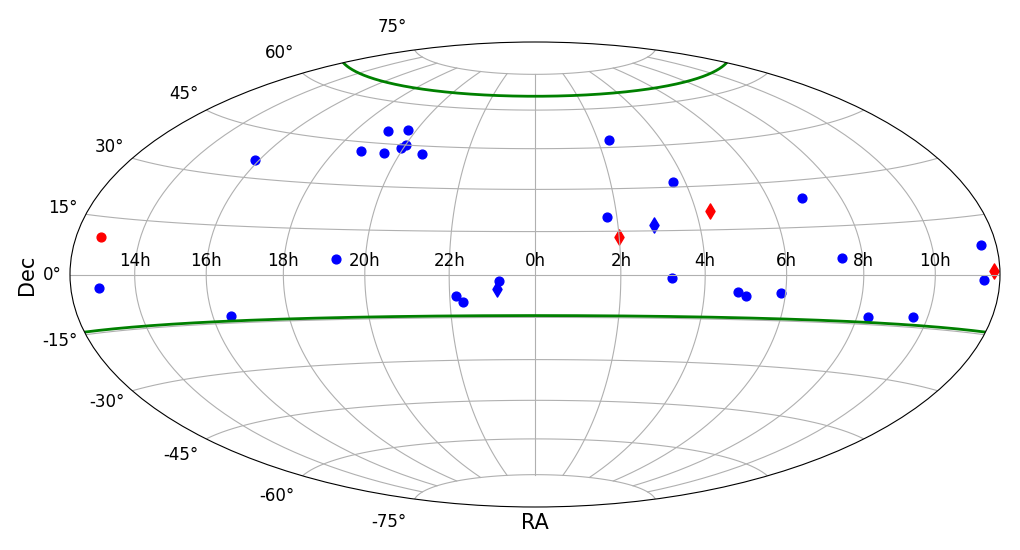}
}
\subfigure[]{
\includegraphics[width=0.48\linewidth]{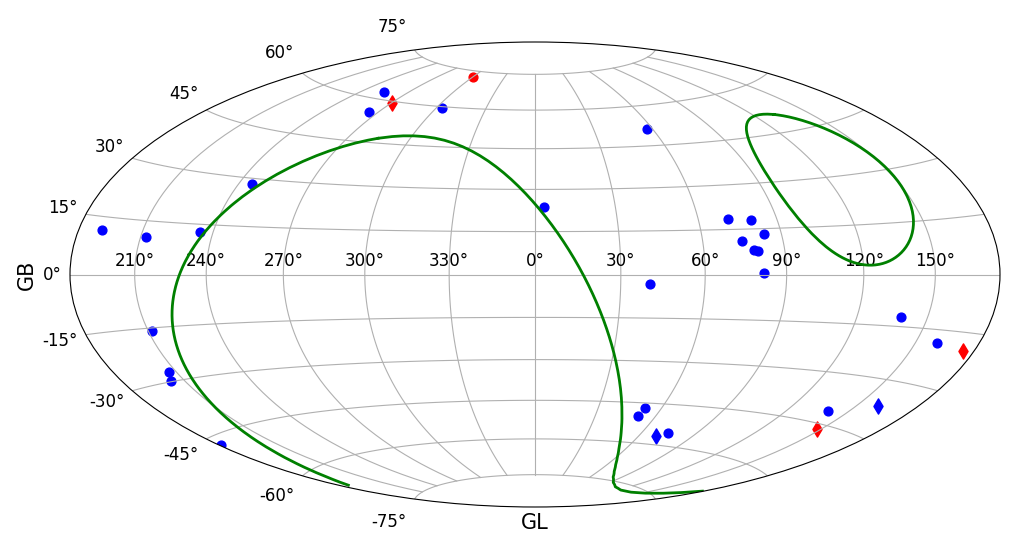}
}
\caption{33 observed targets in equatorial and galactic coordinates. The observable sky of FAST is the region between the green curves. Sources with planets in the habitable zones are denoted by blue symbols. Sources in the Earth transit zone are denoted by diamonds.}
\label{fig:coordinate}
\end{figure}

To improve the theoretical possibility of discovering ETI signals, the observed targets are selected considering habitability, celestial position, and distance, resulting in 29 systems hosting planets in their habitable zones \citep{1993Icar..101..108K,2002IJAsB...1...61W,2013ApJ...765..131K} and 5 systems in the Earth transit zone \citep{2021Natur.594..505K}. The information and coordinates of the sources are shown in Table \ref{tab:targets} and Figure \ref{fig:coordinate}, respectively. Namely, we observe worlds that resemble ours and worlds that can see us. 

Habitable zone is traditionally defined as the circumstellar region in which a terrestrial-mass planet with a CO$_{2}$–H$_{2}$O–N$_{2}$ atmosphere can sustain liquid water on its surface \citep{1993Icar..101..108K}. Whether a planet is located in the habitable zone is determined by the insolation flux it receives, defined as:
\begin{equation}
    F = \left(\frac{R_{\star}}{R_{\odot}}\right)^{2}\left(\frac{T_{\star}}{T_{\odot}}\right)^{4}\left(\frac{a_{\oplus}}{a_{\rm p}}\right)^{2},
\end{equation}
where $R_{\star}$ is the stellar radius, $T_{\star}$ is the stellar effective temperature, and $a_{\rm p}$ is the semi-major axis of the planet's orbit \citep{2014PNAS..11112647B}. 
According to the results of \citet{2013ApJ...765..131K}, the insolation fluxes at the ``recent Venus'' boundary and the ``early Mars'' boundary are respectively calculated by
\begin{equation}
\begin{aligned}
    F_{\text{RV}} = ~&1.7763 + 1.4335 \times 10^{-4} \Delta T + 3.3954 \times 10^{-9} \Delta T^{2}\\
    &- 7.6364 \times 10^{-12} \Delta T^{3} - 1.1950 \times 10^{-15} \Delta T^{4},
\end{aligned}
\end{equation}
\begin{equation}
\begin{aligned}
    F_{\text{EM}} = ~&0.3207 + 5.4471 \times 10^{-4} \Delta T + 1.5275 \times 10^{-9} \Delta T^{2}\\
    &- 2.1709 \times 10^{-12} \Delta T^{3} - 3.8282 \times 10^{-16} \Delta T^{4},
\end{aligned}
\end{equation}
where $\Delta T = T_{\text{eff}\star} - T_{\text{eff}\odot}$. Referring to the data up to October 2020 from the NASA Exoplanet Archive\footnote{\url{https://exoplanetarchive.ipac.caltech.edu}} \citep{https://doi.org/10.26133/nea1}, we select 172 planets around 161 FAST-observable host stars within the optimistic habitable zone (between the ``recent Venus'' and ``early Mars'' boundaries).

Since we cannot observe all the 161 sources within our restricted time, we prioritize them based on two factors. One is the distances, and the other is the sizes of the planets (radii or masses), because the definition of habitable zone is only based on rocky planets, and the size of a planet indicates its surface state. 
We select 32 possible rocky planets orbiting 23 host stars among the 172 planets. The rest of the planets are more likely to have icy or gaseous envelopes. However, compared to the ``icy or gaseous'' planets, the distances to most of the ``rocky'' planets are unfortunately very far. Finally, taking both the distances and the sizes of the planets into account, we observe 12 stars hosting ``rocky'' planets within 220 pc and 17 stars hosting ``icy or gaseous'' planets within 40 pc.

The Earth transit zone is a region where observers beyond the solar system can see the Earth transit the Sun. According the results of \citet{2021Natur.594..505K}, there are 7 known exoplanet systems within 100 pc that are located in the past, present, or future Earth transit zone. Apart from 2 systems included in our habitable zone targets and 2 systems that are not observable by FAST, we observe 3 Earth transit zone systems without known planets in their habitable zones, Ross 128, GJ 9066, and K2-155.

\begin{longtable}{l l c c c c c}%
\caption{Targets}
\label{tab:targets}
\\
\hline%
\hline%
Star Name & Gaia EDR3 ID & Epoch & RA & Dec & Distance (pc) & Observation Date\\
\hline%
\endhead%
\hline%
\endfoot%
\hline%
\endlastfoot%
HD 69830 & 5726982995343100928 & 2016 & 08:18:24.25 & $-$12:38:11.6 & 12.58 & 2021-04-19\\
HD 82943 & 5691782130578684544 & 2016 & 09:34:50.74 & $-$12:07:49.2 & 27.69 & 2021-04-19\\
GJ 96 & 354077348697687424 & 2016 & 02:22:14.99 & $+$47:52:48.8 & 11.95 & 2021-04-20\\
HD 19994 & 3265335443260522112 & 2016 & 03:12:46.64 & $-$01:11:47.1 & 22.88 & 2021-04-20\\
HD 30562 & 3188395880157173120 & 2016 & 04:48:36.72 & $-$05:40:30.5 & 26.14 & 2021-04-20\\
BD-06 1339 & 3022099969137163904 & 2016 & 05:53:00.28 & $-$05:59:47.0 & 20.31 & 2021-04-20\\
HD 180617 & 4293318823182081408 & 2016 & 19:16:54.64 & $+$05:09:46.7 & 5.92 & 2021-04-21\\
HD 218566 & 2638410646295370880 & 2016 & 23:09:11.40 & $-$02:15:40.2 & 28.82 & 2021-04-21\\
Teegarden's Star & 35227046884571776 & 2016 & 02:53:04.71 & $+$16:51:51.7 & 3.83 & 2021-05-25\\
GJ 3323 & 3187115498866675456 & 2016 & 05:01:56.83 & $-$06:56:54.9 & 5.37 & 2021-05-25\\
GJ 273 & 3139847906307949696 & 2016 & 07:27:25.11 & $+$05:12:33.8 & 3.79 & 2021-05-25\\
16 Cyg B & 2135550755683407232 & 2016 & 19:41:51.75 & $+$50:31:00.5 & 21.13 & 2021-05-26\\
HD 210277 & 2619706544757416192 & 2016 & 22:09:29.96 & $-$07:33:02.4 & 21.34 & 2021-05-26\\
TRAPPIST-1 & 2635476908753563008 & 2016 & 23:06:30.37 & $-$05:02:36.7 & 12.47 & 2021-05-26\\
Wolf 1061 & 4330690742322011520 & 2016 & 16:30:17.96 & $-$12:40:04.3 & 4.31 & 2021-06-06\\
K2-72 & 2615653023342995584 & 2016 & 22:18:29.47 & $-$09:36:43.2 & 66.51 & 2021-06-21\\
HD 10697 & 95652018353917056 & 2016 & 01:44:55.78 & $+$20:04:57.7 & 33.16 & 2021-06-21\\
HD 22781 & 217334764042444288 & 2016 & 03:40:49.58 & $+$31:49:33.1 & 32.56 & 2021-06-21\\
HD 50554 & 3380479015342121600 & 2016 & 06:54:42.78 & $+$24:14:42.5 & 31.07 & 2021-06-21\\
K2-3 & 3796690380302214272 & 2016 & 11:29:20.49 & $-$01:27:18.5 & 43.98 & 2021-06-21\\
HD 111998 & 3679242575447826432 & 2016 & 12:53:10.88 & $-$03:33:11.2 & 33.44 & 2021-06-21\\
kap CrB & 1372702716380418688 & 2016 & 15:51:13.92 & $+$35:39:21.0 & 29.99 & 2021-06-21\\
Kepler-1649 & 2125699062780742016 & 2016 & 19:30:00.71 & $+$41:49:47.9 & 92.76 & 2021-06-22\\
Kepler-186 & 2079000330051813504 & 2016 & 19:54:36.66 & $+$43:57:18.0 & 177.51 & 2021-06-22\\
Kepler-560 & 2082162147537254400 & 2016 & 20:00:49.60 & $+$45:01:05.7 & 109.43 & 2021-06-22\\
HD 197037 & 2066437688140905600 & 2016 & 20:39:32.87 & $+$42:14:51.2 & 33.08 & 2021-06-22\\
Kepler-296 & 2132069633148965888 & 2016 & 19:06:09.61 & $+$49:26:14.1 & 219.60 & 2021-06-22\\
Kepler-438 & 2104675781979819776 & 2016 & 18:46:34.97 & $+$41:57:03.8 & 179.88 & 2021-06-29\\
GJ 9066 & 76868614540049408 & 2016 & 02:00:14.16 & $+$13:02:38.7 & 4.47 & 2021-09-10\\
K2-155 & 145333927996558976 & 2016 & 04:21:52.71 & $+$21:21:11.7 & 72.93 & 2021-09-10\\
K2-18 & 3910747531814692736 & 2016 & 11:30:14.43 & $+$07:35:16.1 & 38.10 & 2021-09-10\\
Ross 128 & 3796072592206250624 & 2016 & 11:47:45.05 & $+$00:47:56.8 & 3.37 & 2021-09-10\\
Gliese 486 & 3735000631158990976 & 2016 & 12:47:55.53 & $+$09:44:57.7 & 8.08 & 2021-09-21
\end{longtable}

\subsection{Strategy} \label{subsec:strategy}

The basic principle of the on-off strategy is that an extraterrestrial signal from the target can only be present in the on-observation, while RFI entering the side lobes can be detected in the off-observation \citep{2017ApJ...849..104E}. Corresponding to the on-off strategy, the beam tracking the target (target beam) in the MBCM strategy serves as the on-observation, and some of the other beams (reference beams) serve as the off-observations. Therefore, the fundamental requirement of rejecting RFI by MBCM is that an extraterrestrial signal received by the target beam is not detectable by the reference beams. The layout of the 19 beams on the FAST L-band receiver is shown in Figure \ref{fig:beam}. The actually measured FWHM beamwidth of each beam increases with its distance to Beam 1 (central beam), and decreases with frequency. According to the data in Table 2 of \citet{2020RAA....20...64J}, the average FWHM beamwidth of the six outermost beams (Beams 8, 10, 12, 14, 16 and 18) at 1060 MHz is $3.51^{\prime}$. Because the angular distance between these six beams and Beam 1 is $\sim 11.6^{\prime}$, exceeding 3 FWHM of the six beams, using them as the reference beams for Beam 1 satisfies the criterion used by \citet{2020AJ....159...86P} for the on-off strategy on Parkes. This means that although bright ETI signals from the target may cover several of the six adjacent to Beam 1 (Beams 2, 3, 4, 5, 6 and 7), they are unlikely to be detected by the six outermost beams. 

\begin{figure}[htbp]
\centering
\includegraphics[width=0.5\textwidth]{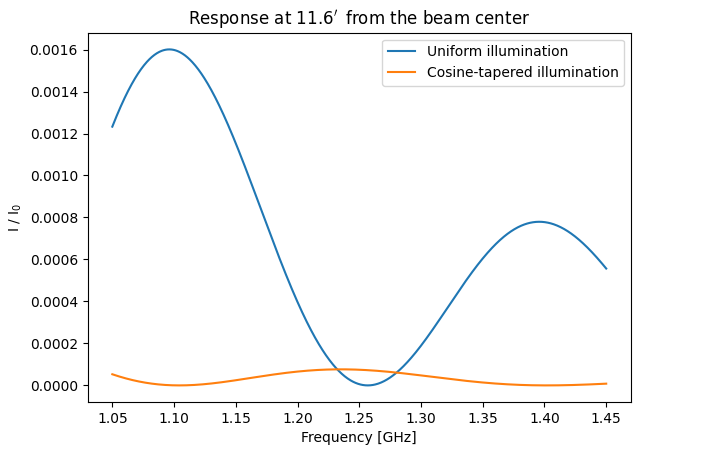}
\caption{\label{fig:response} The theoretical beam response at 11.6$^{\prime}$ from the beam center for uniform illumination and cosine-tapered illumination for the FAST L-band 19-beam receiver.}
\end{figure}

\begin{figure}[htbp]
\centering
\subfigure[]{
\includegraphics[width=0.25\linewidth]{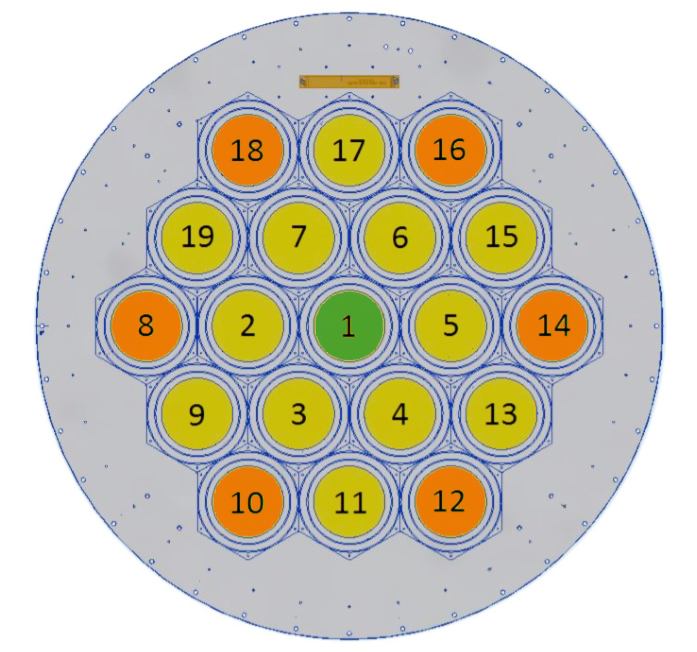}
}
\subfigure[]{
\includegraphics[width=0.25\linewidth]{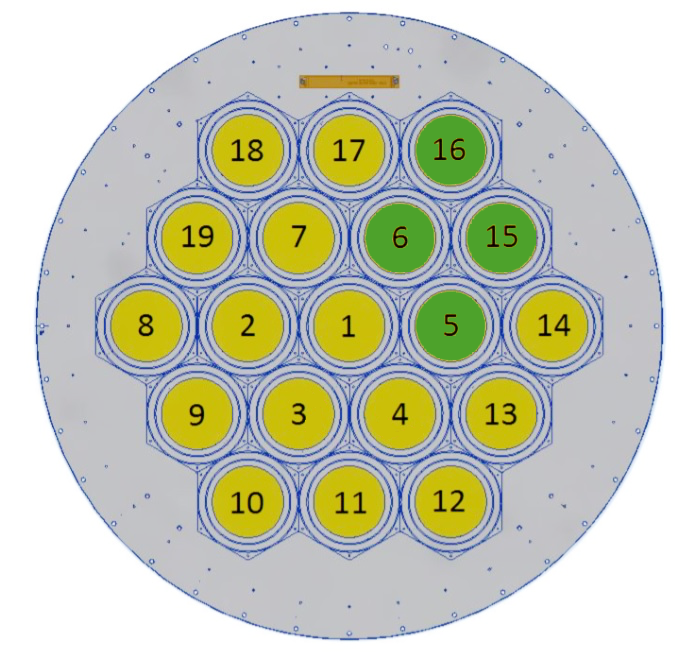}
}
\subfigure[]{
\includegraphics[width=0.25\linewidth]{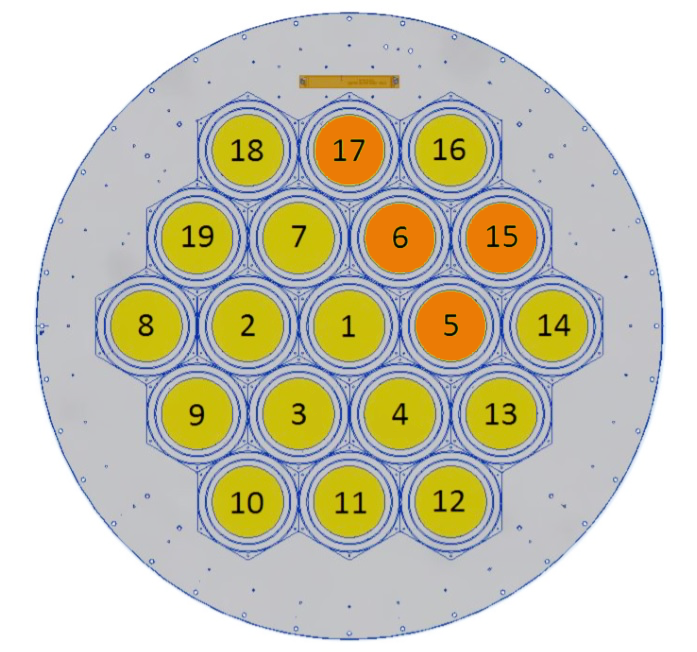}
}
\caption{Schematics of the MBCM strategy. (a) In MBCM targeted searches, an ETI signal detected by Beam 1 cannot appear in the 6 outermost beams, otherwise it is RFI. (b) An example of permitted signals in MBCM blind searches. (c) An example of forbidden signals in MBCM blind searches. Beams 5, 6 and 17 are arranged in a line, thus an extraterrestrial signal cannot cover them simultaneously.}
\label{fig:beam}
\end{figure}

To verify this more quantitatively, we can measure the response of an outermost beam at the center of Beam 1, which is $11.6^{\prime}$ from the center of the outermost beam. Figure 6 and 7 of \citet{2020RAA....20...64J} show that this response is very small compared to the response at the beam center, but the specific value of this response is not given. The measurement of the FWHM beamwidths of the FAST L-band 19-beam receiver indicates that the illumination pattern of FAST is between the uniform illumination and cosine-tapered illumination \citep{2019SCPMA..6259502J}. Given the 300-meter illuminated aperture of FAST, the variation of the response at $11.6^{\prime}$ from the beam center with frequency for the uniform illumination and cosine-tapered illumination is shown in Figure \ref{fig:response}. Theoretically, the actual response should be roughly between the two curves, which should be very small. Hence, in order to identify whether the signals received by Beam 1 are ETI technosignatures or RFI, we use the six outermost beams as the reference beams (Figure \ref{fig:beam}(a)).

During an observation, we use all 19 beams on the L-band receiver to record data simultaneously, and Beam 1 keeps tracking the target in the process. Each observation lasts for 20 min (except for HD 111998, observed for $\sim$4 min). Beams other than the seven mentioned above are used for signal identification in this work, which will be discussed below. We refer to the narrow-band signals detected above the signal-to-noise ratio (S/N) threshold by any of the 19 beams as ``hits''. Among the hits detected by Beam 1, we refer to those not detected by any of the reference beams as ``events''. The signals detected by both beam 1 and any of the reference beams are determined as RFI and rejected directly. Using as many as six reference beams, we are able to remove the vast majority of signals detected by Beam 1 effectively.

Apart from targeted observations, the MBCM strategy can also be applied to blind searches for ETI technosignatures. Our criteria refer to the multi-beam blind search strategy for FRB on FAST \citep{2019SSPMA..49i9508P,2020ApJ...895L...6Z}. We require that the maximum number of covered beams by an extraterrestrial signal is four, and such a signal cannot cover three beams arranged in a line or beams spanning a larger distance. Hence, there are four types of beam coverage patterns for permitted signals: (1) A single beam; (2) two adjacent beams; (3) three beams adjacent to each other forming an equilateral triangle; and (4) four beams forming a compact rhombus as shown in Figure \ref{fig:beam}(b). Signals that violate the above criteria are determined to be RFI, such as a signal with the beam coverage shown in Figure \ref{fig:beam}(c), where Beams 5, 6 and 17 are arranged in a line. In this work, we are only concerned with ETI signals from the exoplanet targets, so this blind search strategy will be practiced in the future.

\section{Data Analysis} \label{sec:DataAnalysis}

We record our data using the spectral line backend with the L-band 19-beam receiver across $1.0 - 1.5$ GHz. The frequency resolution of the spectra is $\sim$7.5 Hz and the integration time of each spectrum is 10 seconds. Each FITS file contains four polarization channels of two spectra recorded by one beam, and the total volume of our data is 66.5 TB (including calibration observations). The FITS files of one beam observing one target are concatenated and converted into two Filterbank files (XX and YY), a data format accessible to the Blimpy Python package \citep{2019JOSS....4.1554P}.

\subsection{Signal Search}

\begin{figure}[htbp]
\centering
\includegraphics[width=0.3\textwidth]{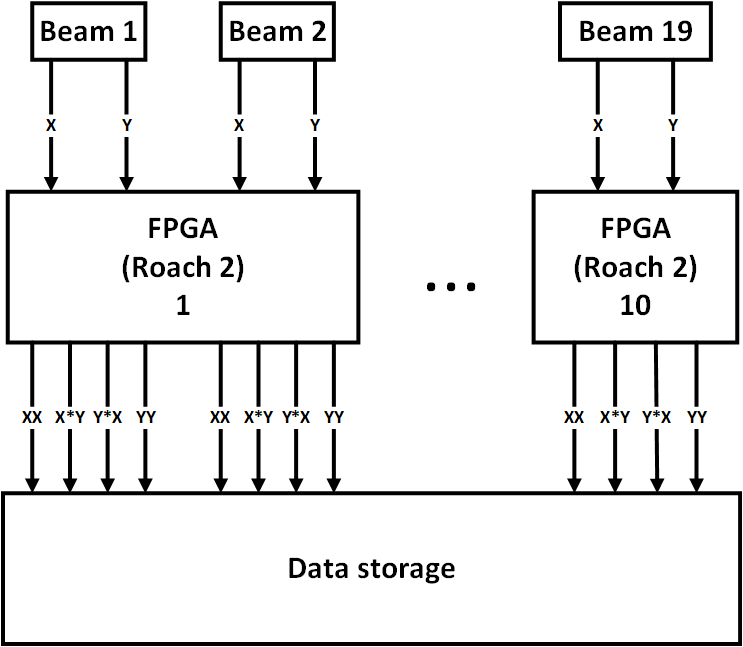}
\caption{\label{fig:backend} A schematic of the FAST multi-beam digital backend for the L-band 19-beam receiver. Every two beams share one Roach 2 FPGA board, except for Beam 19. Before processed by Roach 2, the signals in polarization directions X and Y are transmitted and sampled separately. After self-correlation and cross-correlation of the data of two polarization directions, the data output by the FPGA for one beam consists of 4 polarization channels: XX, YY, X*Y and Y*X.}
\end{figure}

Most targeted SETI in recent years search only Stokes I data for narrow-band drifting signals \citep{2017AJ....153..110G,2020AJ....159...86P,2020AJ....160...29S,2021NatAs.tmp..203S}. The observation data output by the FAST multi-beam digital backend consists of four polarization channels: XX, YY, X*Y and Y*X \citep{2020RAA....20...64J,2021RAA....21..107H}. They are derived by self-correlation and cross-correlation of the data in two orthogonal linear polarization directions X and Y, and this conversion is done on the Roach 2 field programmable gate array (FPGA) board in the FAST multi-beam digital backend (Figure \ref{fig:backend}). We search the intensity data of two orthogonal linear polarization directions (XX and YY) separately for narrow-band signals. 

TurboSETI \citep{2017ApJ...849..104E,2019ascl.soft06006E} is a Python/Cython package using the tree search algorithm \citep{2013ApJ...767...94S} to search for narrow-band signals. It breaks the limitation of the time-frequency resolution by shifting arrays, thus  is able to search for signals with arbitrarily large drift rates. Two essential parameters required by turboSETI are an S/N threshold and a maximum drift rate (MDR), which allow turboSETI to search for narrow-band signals above the S/N threshold within $\pm$MDR. Referring to previous targeted SETI studies using turboSETI \citep{2017ApJ...849..104E,2020AJ....159...86P,2020AJ....160...29S,2021AJ....161..286T,2021AJ....162...33G,2021NatAs.tmp..203S}, we adopt the most commonly used values for these parameters, setting the S/N threshold to 10 and the MDR to 4 Hz/s. Too high or too low S/N threshold will lead to too few or too many events. Since the drift rate of a signal is proportional to its original frequency, the suitable MDR depends on the observing frequency band, and 4 Hz/s is large enough for FAST L-band. TurboSETI outputs the best-fit frequencies, drift rates, and S/Ns of the hits detected by each beam to a DAT file.

We use the find\_event\_pipeline of turboSETI to select events and reject obvious RFI. The RFI identification criteria in the original code applies to to a standard on-off strategy, where on- and off- observations are taken in turn, thus the same RFI signal appears at different frequencies in a series of observations due to the Doppler drift. Because the MBCM strategy observes a target and reference locations simultaneously, an RFI signal is expected to appear at the same frequency in all beams. We modify the code and set the RFI rejection range as $\pm3 \delta\nu$, where $\delta\nu$ is our frequency resolution. That is, if a hit in Beam 1 is accompanied by any hit within this range in the reference beams, the hit will be determined as RFI. TurboSETI outputs the selected events of each source to a CSV file. Since the effective bandwidth of the L-band receiver is $1.05 - 1.45$ GHz \citep{2011IJMPD..20..989N}, we discard events detected within the two unusable 50 MHz-wide ends of the band after event finding by turboSETI.

\subsection{De-drifting}

Though turboSETI provides a first-order best-fit drift rate by the tree search algorithm for every hit, we can determine the drift rate to higher precision by a de-drifting algorithm to further analyze the characteristics of particular signals.

Around the first-order drift rate estimated by turboSETI, we set a drift rate test range and step size for a specific signal. Then we intercept a time-frequency-power array around the frequency of the signal. At each drift rate, every time row in the time-frequency-power array is shifted by the corresponding number of pixels, calculated by
\begin{equation}
    \text{number of pixels}=\left[\text{row index}\times \text{drift rate}\times \frac{\delta t}{\delta \nu} \right],
\end{equation}
to form a stair-like array, and this procedure is named ``de-drifting''. Once the array is de-drifted, the spectra in the array are summed over the observation time. From this one-dimensional spectrum, we search for a power maximum in frequency space, and each drift rate corresponds to such a maximum. The drift rate that yields the highest power maximum is the best-fit one. The precision of the drift rate obtained by this algorithm is higher than that fitted by the tree search algorithm. Using the de-drifted time-frequency-power array, the frequencies, drift rates, and S/Ns of weak signals that are visible on the waterfall plots can also be determined, which may fail to reach the S/N threshold and are thus neglected by turboSETI.


\section{Results} \label{sec:Results}

\subsection{Results of Signal Search}

We find 1,309,503 hits of XX and 1,324,198 hits of YY from all 19 beams. Among the hits of Beam 1, we select 2,013 events of XX and 2,064 events of YY, excluding 97.0\% and 96.9\% of the hits detected by Beam 1, respectively.

\begin{figure}[htbp]
\centering
\subfigure[]{
\includegraphics[width=0.8\linewidth]{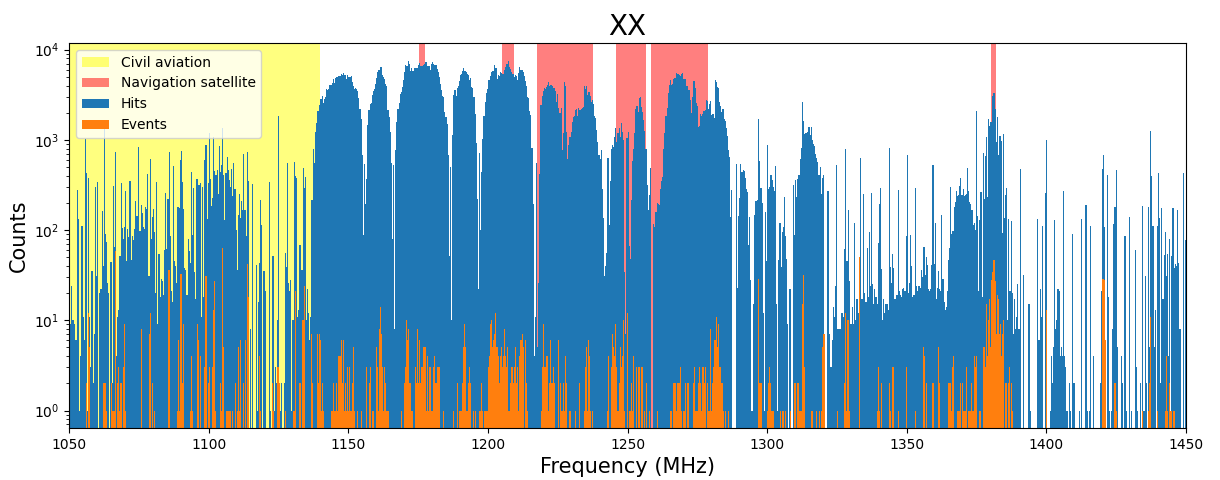}
}

\subfigure[]{
\includegraphics[width=0.4\linewidth]{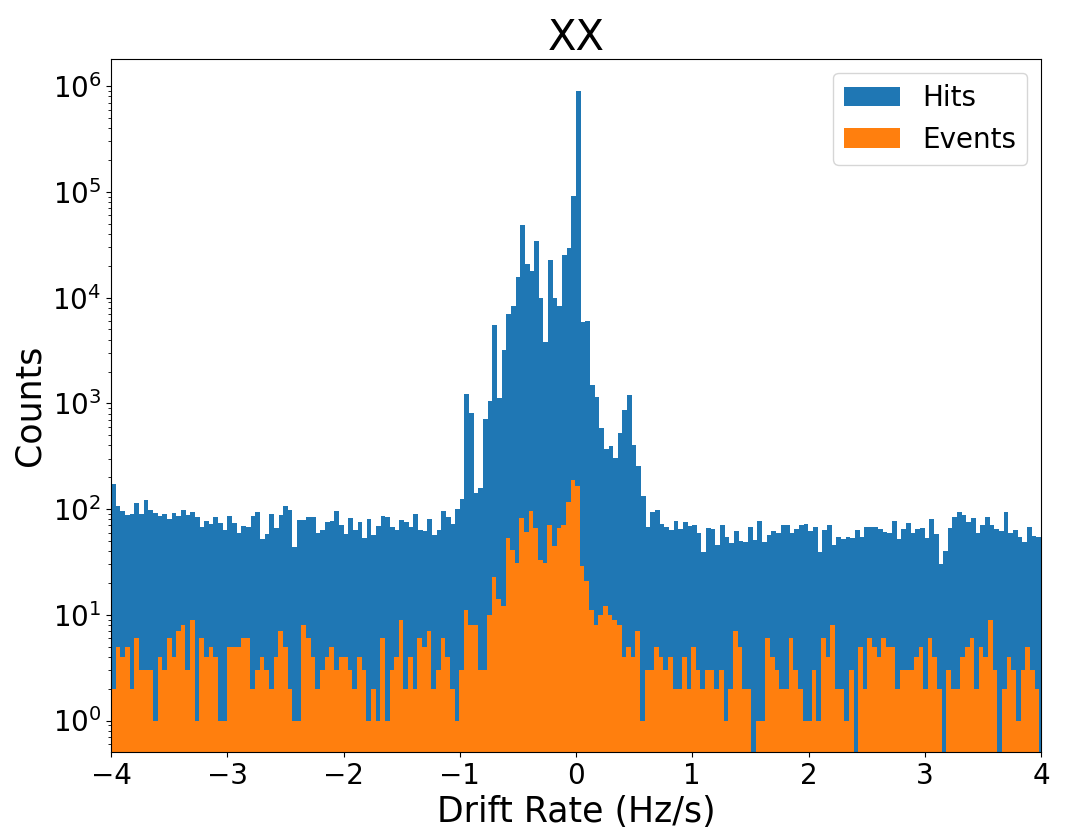}
}
\subfigure[]{
\includegraphics[width=0.4\linewidth]{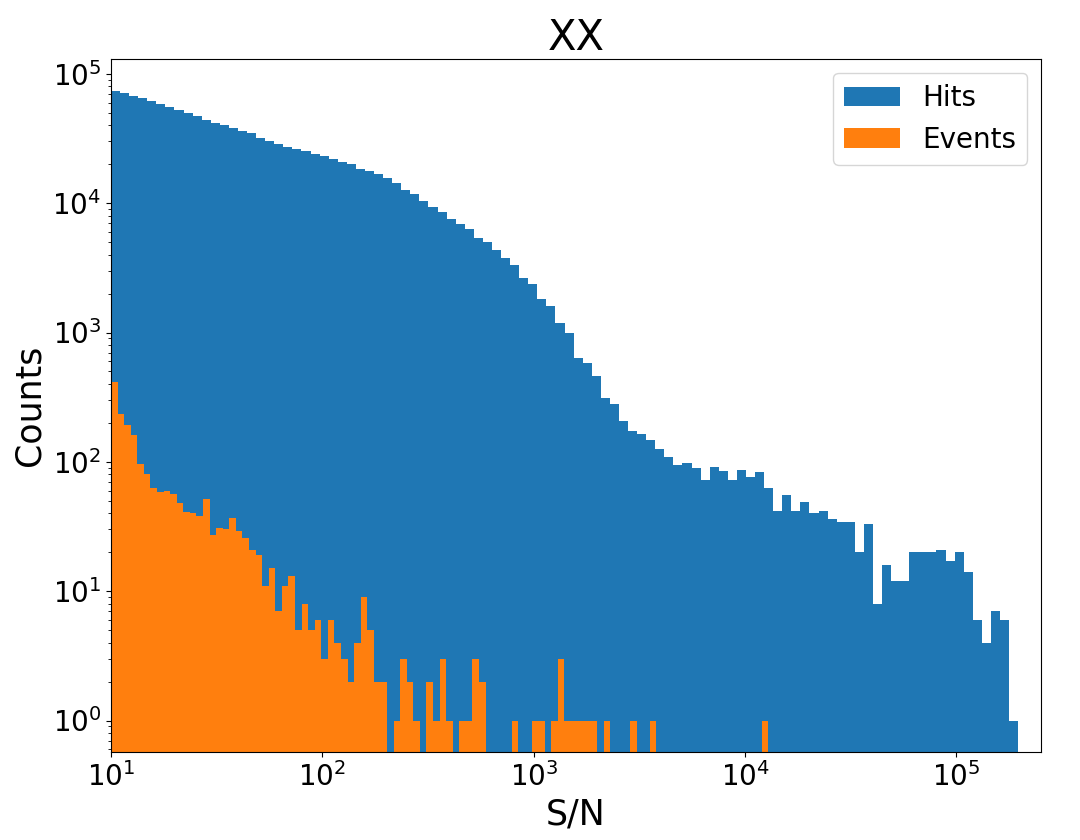}
}
\caption{Histograms of hits and events for XX as functions of frequency, drift rate and S/N. Frequency bands of catalogued interference sources are displayed on the frequency panel. The blue bars show the distributions of the hits detected by all 19 beams. The orange bars show the distributions of the events detected by Beam 1.}
\label{fig:statistics1}
\end{figure}

\begin{figure}[htbp]
\centering
\subfigure[]{
\includegraphics[width=0.8\linewidth]{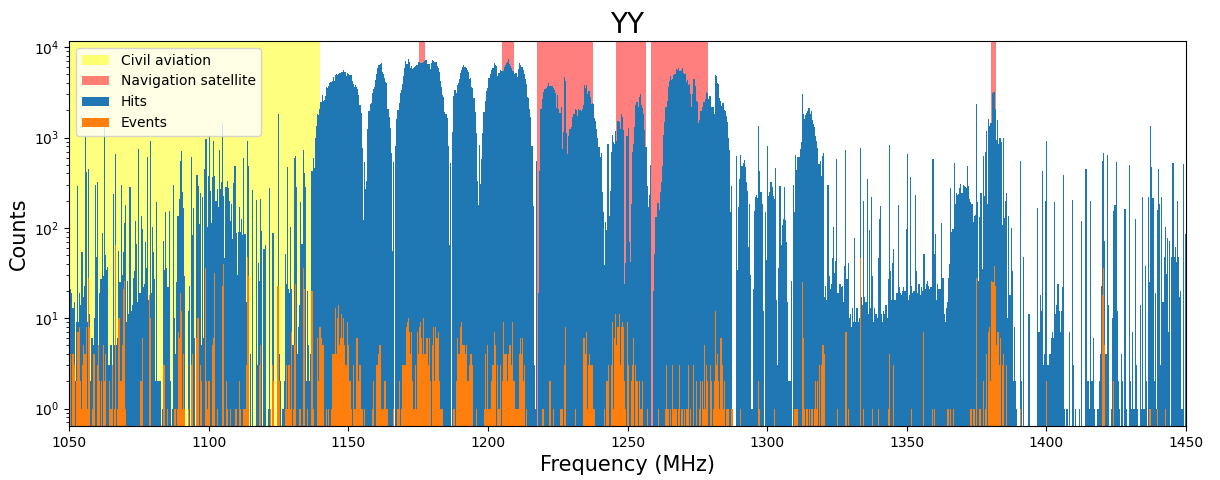}
}

\subfigure[]{
\includegraphics[width=0.4\linewidth]{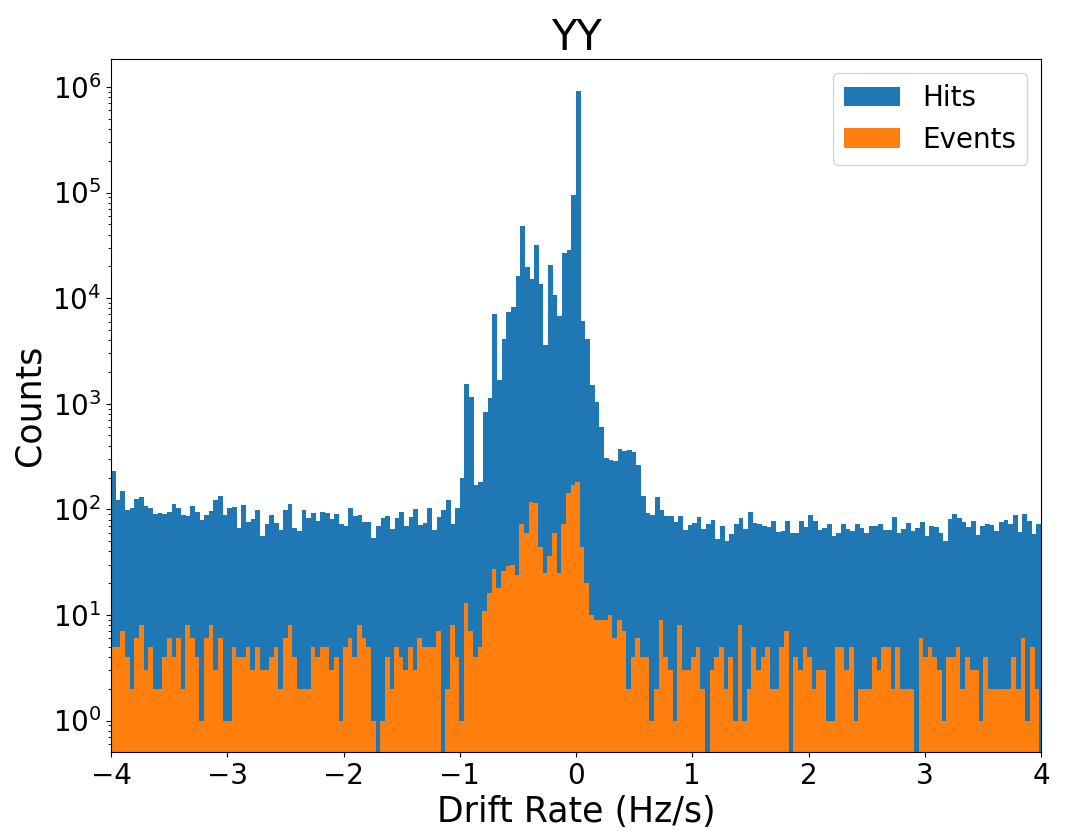}
}
\subfigure[]{
\includegraphics[width=0.4\linewidth]{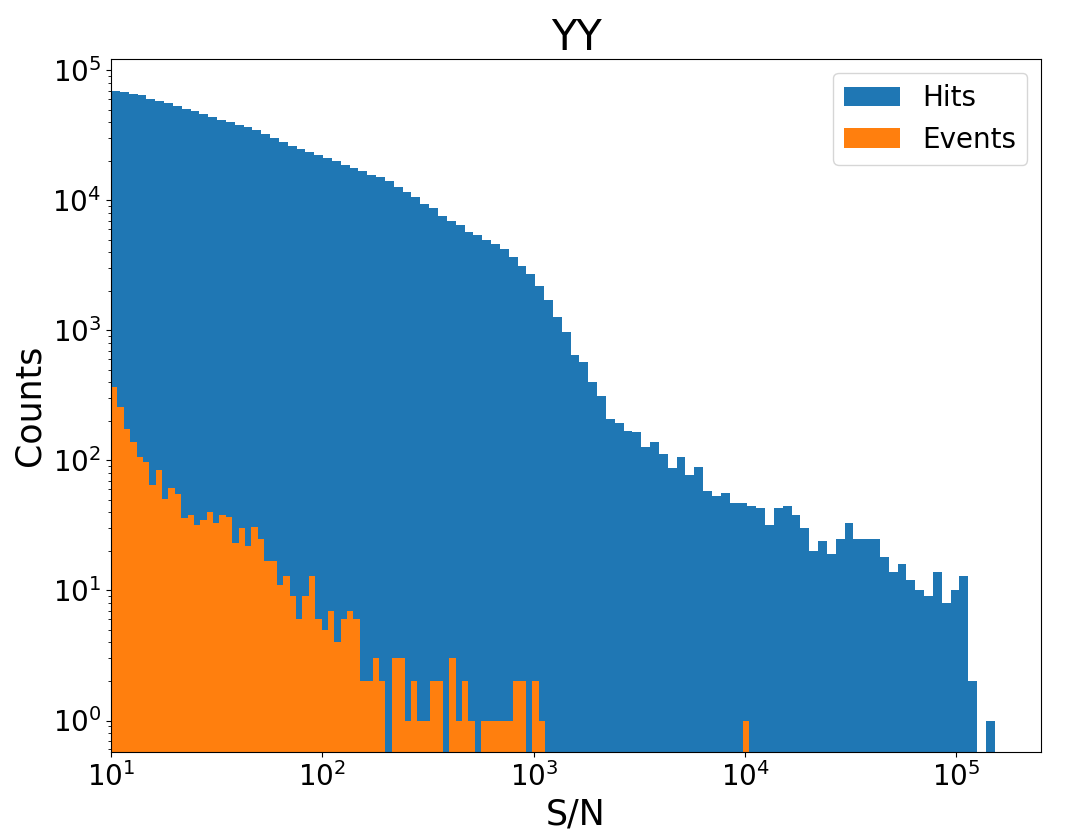}
}
\caption{Histograms of hits and events for YY as functions of frequency, drift rate and S/N.}
\label{fig:statistics2}
\end{figure}

The distributions of frequency, drift rate, and S/N for the hits detected by all 19 beams and the events detected by Beam 1 are shown in Figure \ref{fig:statistics1} and \ref{fig:statistics2}, and each figure displays one polarization direction. The general distribution trend of each parameter does not vary apparently with polarization direction, implying that most hits and events are not significantly polarized. According to the results of RFI environment tests at the FAST site, there are two major types of RFI sources within the $1.05-1.45$ GHz frequency band, civil aviations and navigation satellites \citep{2021RAA....21...18W}. We find only a small fraction (4.2\% and 3.5\% for XX and YY, respectively) of the hits falling within the civil aviation band ($1030 - 1140$ MHz). In contrast, hits within the navigation satellite bands ($1176.45 \pm 1.023$ MHz, $1207.14 \pm 2.046$ MHz, $1227.6 \pm 10$ MHz, $1246.0 - 1256.5$ MHz, $1268.52 \pm 10.23$ MHz, $1381.05 \pm 1.023$ MHz) account for a considerable proportion of the hits (31.5\% and 31.6\% for XX and YY, respectively), implying that they are a major RFI source. The proportions of the hits with positive, zero, and negative drift rates are 5.6\%, 65.8\% and 28.6\% for XX, and 5.4\%, 66.1\% and 28.5\% for YY. Non-drift hits are in the majority as expected, since ground-based RFI sources are mostly stationary. The bias towards negative drift rate results from the downward relative acceleration vectors of non-geosynchronous satellites. The majority of the events have low S/N because weak RFI signals are likely to be detected below the S/N threshold by the reference beams but above the S/N threshold by Beam 1 accidentally, thus passing our event selection.

\subsection{Signal Identification}

\begin{figure}[htbp]
\centering
\subfigure[]{
\includegraphics[width=0.4\linewidth]{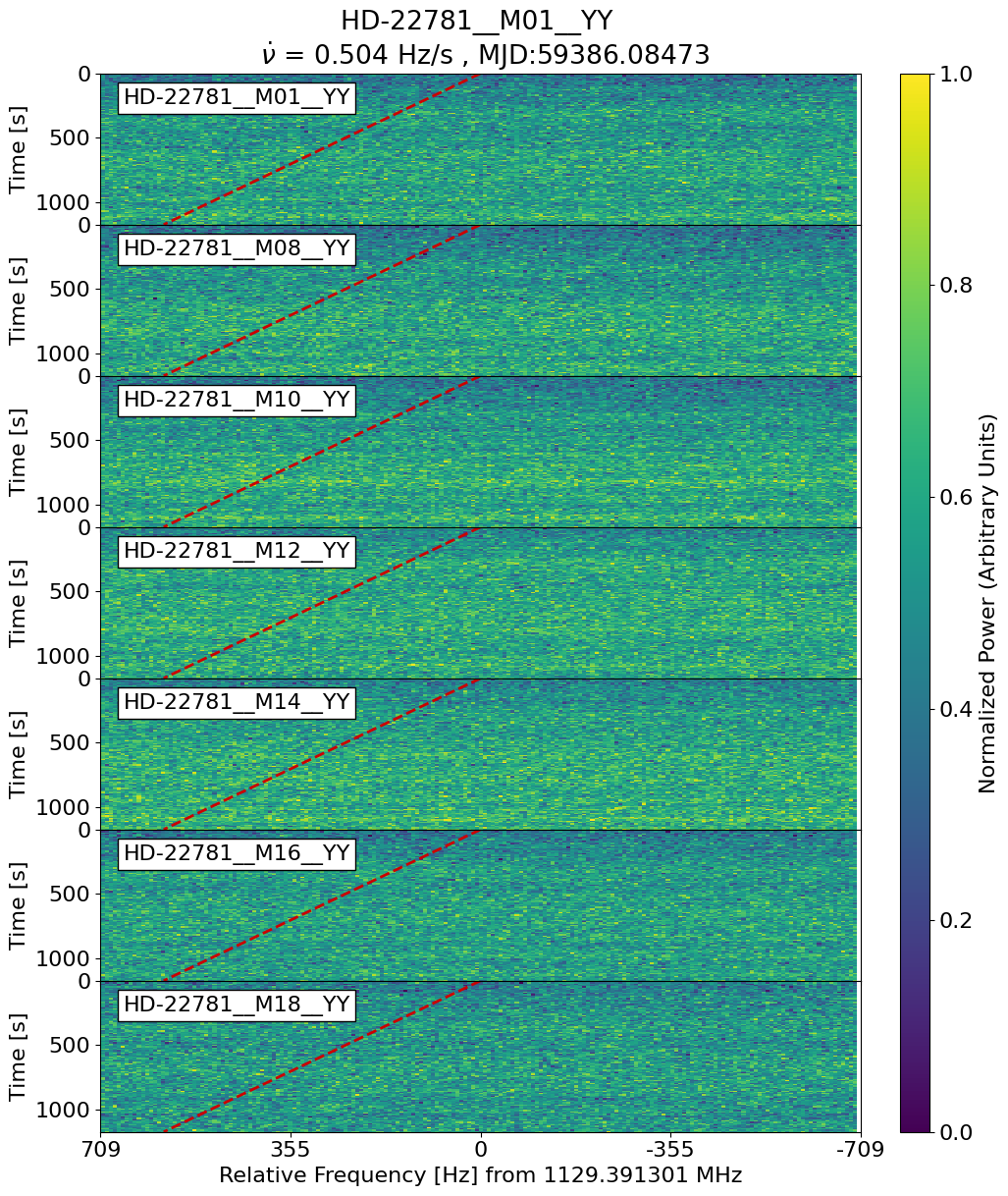}
}
\subfigure[]{
\includegraphics[width=0.4\linewidth]{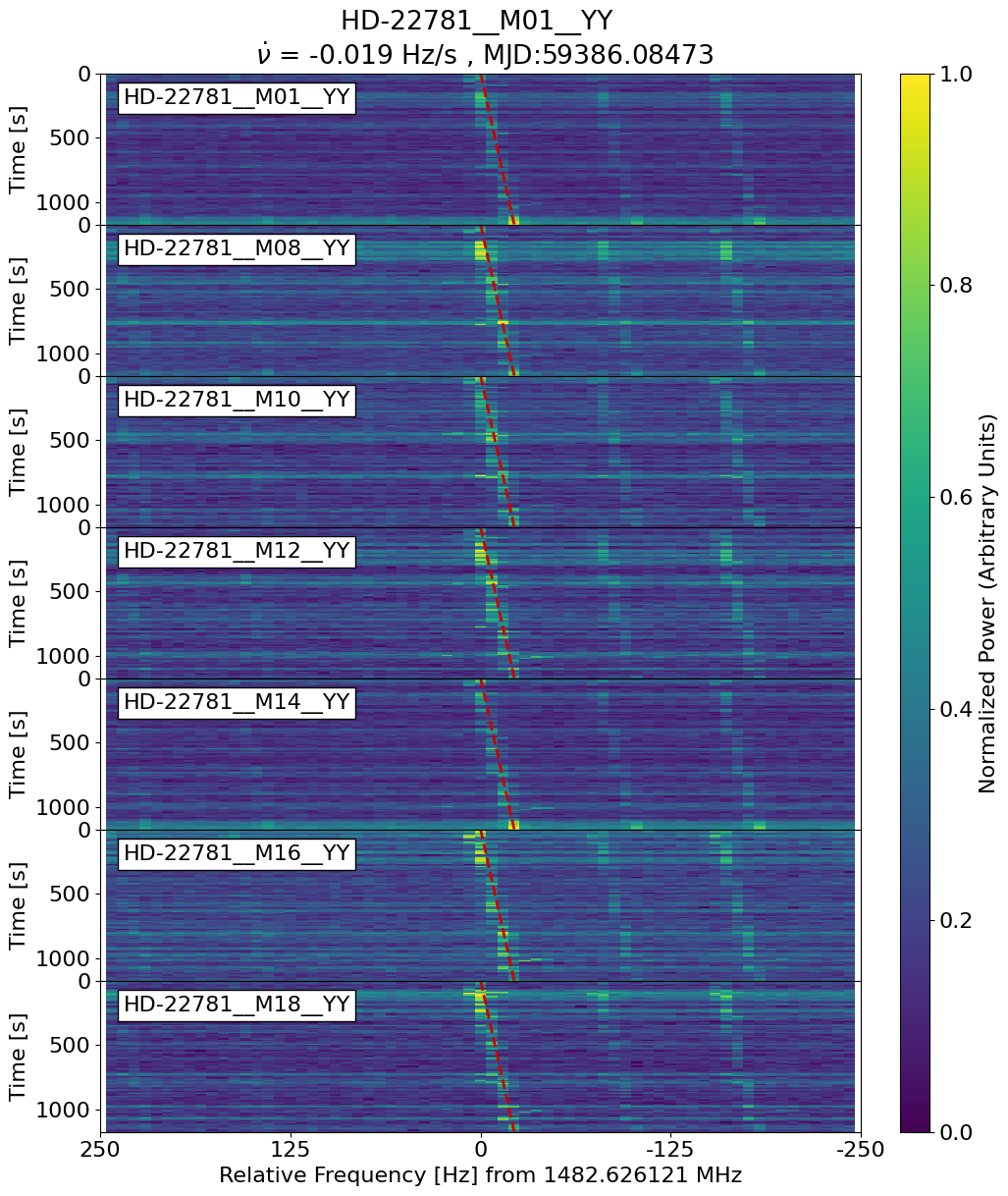}
}
\caption{Examples of false positives. (a) An event visible in the reference beams. (b) An event invisible in Beam 1. The red dashed lines show the frequencies and drift rates of the hits in Beam 1 fitted by turboSETI.}
\label{fig:false_positives}
\end{figure}

All the events selected by the program are re-examined by visual inspection of the dynamic time–frequency spectra (waterfall plots). We find that most events ($\sim$93\%) are apparent false positives and can be excluded directly. The false positives can be classified into two forms: (1) there is no visible signal in Beam 1, but the program detects a hit (Figure \ref{fig:false_positives}(a)); (2) there are visible signals in the reference beams at an approximate frequency to the hit in Beam 1, but the program judges it as an event (Figure \ref{fig:false_positives}(b)). For the former case, we verify that there is indeed no signal in Beam 1 by integrating the spectra over time after shifting them by the drift rate (de-drifting). The latter case can be caused by two reasons: (1) the signal is so weak that it fails to reach the S/N threshold in the reference beams; (2) the hits in the reference beams are fitted at slightly different frequencies from the hit in Beam 1, and the differences exceed the RFI rejection range we set in this work.

\begin{figure}[htbp]
\centering
\includegraphics[width=0.8\textwidth]{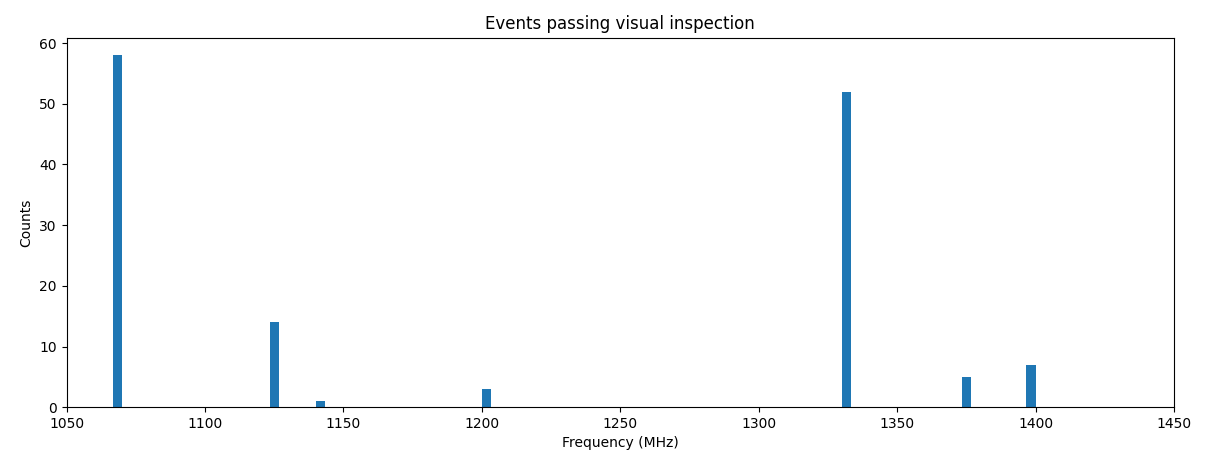}
\caption{\label{fig:events} The frequency distribution of the 140 events passing visual inspection.}
\end{figure}

After discarding the false positives, the remaining events of XX and YY are combined to yield 140 events. The frequency distribution of these 140 events is displayed in Figure \ref{fig:events}. We find that except for an event detected at 1140.604 MHz, the frequencies of these events are concentrated at six values. Here we list the numbers of events, mean values and standard deviations of the six clusters: (1) 58 events around $1066.6658 \pm 0.0131$ MHz; (2) 14 events around $1124.9967 \pm 0.0001$ MHz; (3) 3 events around $1200.0104 \pm 0.0001$ MHz; (4) 52 events around $1333.3270 \pm 0.0129$ MHz; (5) 5 events around $1375.0009 \pm 0.0049$ MHz; and (6) 7 events around $1400.0207 \pm 0.0003$ MHz. These frequencies can be linearly combined by the nominal frequencies (33.3333 MHz and 125.00 MHz) of the clock oscillators used by the Roach 2 FPGA board, so these events are attributed to the harmonics generated by the clock oscillators. RFI produced by the instruments in the backend rather than entering from the side lobes is called instrumental RFI. 

The hardware used for digital signal sampling in the FAST multi-beam digital backend consists of the Roach 2 FPGA board and KatADC, which are attached to each other and placed in the Roach 2 enclosure. Crystal oscillators of various frequencies are required to provide clock signals for the peripherals of Roach 2. The signals are transmitted to the devices through traces on the circuit board, and these traces act as antennas. Therefore, the space in the enclosure is filled with electromagnetic waves of different frequencies, which can be coupled to the pins of KatADC and amplified by the amplifiers in the sampling frontend. When the intensity reaches a certain degree, they can be sampled and finally become the instrumental RFI we see on the spectra. The nominal frequencies of all crystal oscillators used can be found in the BOM\footnote{\url{https://github.com/ska-sa/roach2_hardware/blob/master/release/rev2/A/BOM/ROACH-2_REV2_BOM.csv}} of Roach 2. The ADC sampling frequency is 1 GHz and the FPGA operating frequency is 250MHz. The linear combinations of all these frequencies are likely to become the frequencies of detected instrumental RFI. In addition, since the Roach 2 enclosures are placed in a computer room full of various electronic devices, where the electromagnetic environment is badly contaminated. It is also possible for strong external electromagnetic interference to leak into the enclosure through photoelectric conversion and other analog channels in front of Roach 2, so as to form a detectable RFI.

A commonality for these harmonics is that they are also present in Beam 2. 
The reason is that in the backend every two beams share one Roach 2 FPGA board (except for Beam 19, see Figure \ref{fig:backend}), so the clock oscillators in the same Roach 2 enclosure can affect the data of both beams simultaneously. The actual frequency of a clock oscillator is susceptible to various factors such as temperature, aging and voltage, thus the frequencies of the harmonics are unstable and can vary in the range of hundreds of Hertz.


\subsection{Identifying a Particular Signal by Polarization}

\begin{figure}[htbp]
\centering
\includegraphics[width=0.85\linewidth]{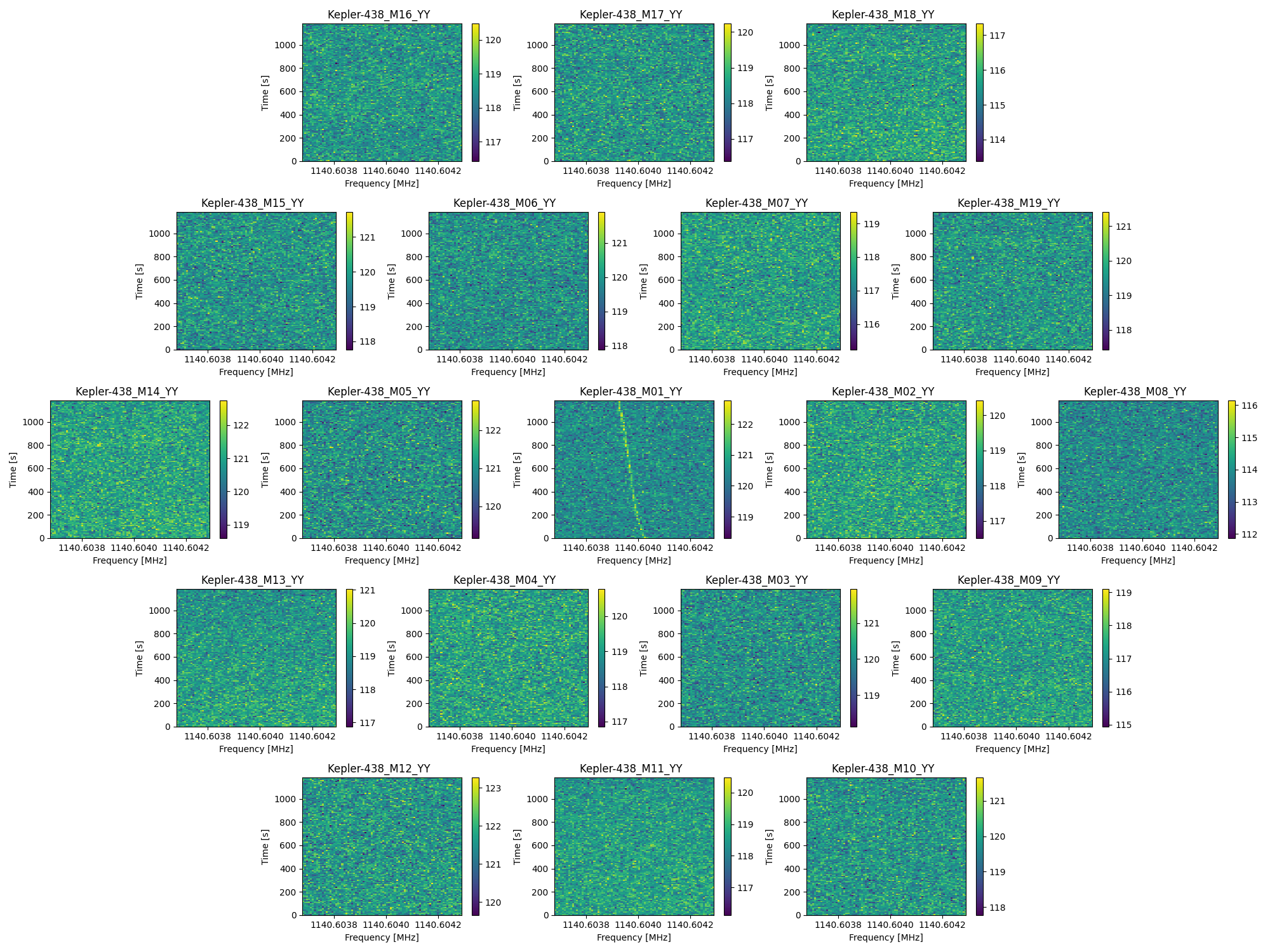}
\caption{A signal at 1140.604 MHz detected during our 20-minute observation towards Kepler-438. This signal is only present in Beam 1, not in any other beam, and is the only signal that satifies this characteristic in our observations.}
\label{fig:candidate}
\end{figure}

So far, we have rejected all the events except the one detected at 1140.604 MHz from the observation towards Kepler-438. This event piqued our interest at first because its bandwidth ($\sim$Hz) and best-fit drift rate ($-0.0678$ Hz/s) are within the reach of a transmitter moving with an exoplanet, and its frequency cannot be explained by known clock oscillator harmonics. In addition, as shown in Figure \ref{fig:candidate}, it is the only event that is present in Beam 1 and not in any other beam, which makes it the most special event in this work. However, based on its polarization characteristics,  we are able to eliminate the possibility of an extraterrestrial origin.

It is possible that ETI signals are linearly polarized. If an ETI signal is not linearly polarized, it will display similar intensities in XX and YY on the spectra, just like the neutral hydrogen line (Figure \ref{fig:HI}(a)). But if an ETI signal is linearly polarized, and its polarization direction coincides with X or Y direction of the feed, the intensity in one polarization channel (XX or YY) will be much stronger than that in the other. In that case, the relative intensity between XX and YY should vary with the parallactic angle of the source, and the variation should be more obvious with a longer observation duration. 

\begin{figure}[htbp]
\centering
\subfigure[]{
\includegraphics[width=0.4\linewidth]{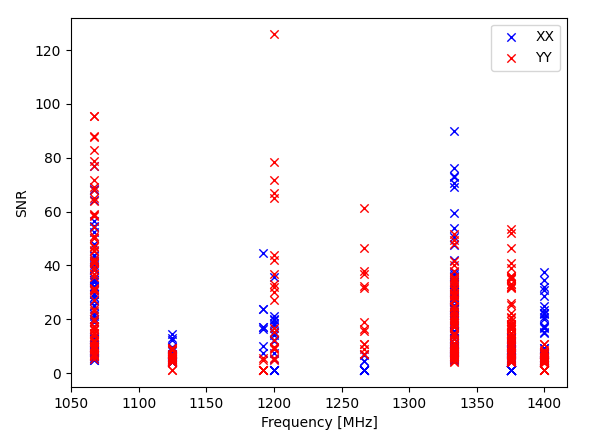}
}
\subfigure[]{
\includegraphics[width=0.4\linewidth]{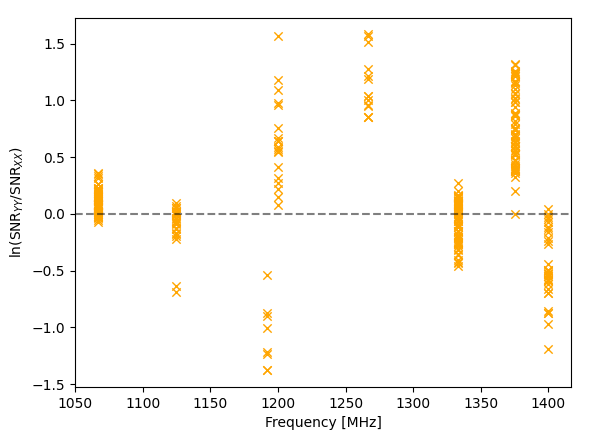}
}
\caption{\label{fig:crystal_snr} The S/N distribution with frequency of clock oscillator harmonics. The statistics here involve 298 clock oscillator harmonics in Beam 1. Some of them are not detected by turboSETI, but are visible on the waterfall plots. If a signal is totally absent from the waterfall plots in one polarization (but present in the other polarization), then the S/N of the signal in that polarization is determined as 1. (a) The S/N distribution in XX and YY. (b) The ratios of S/N of YY to XX in logarithm. The signals above the horizontal dashed line are stronger in YY than XX, otherwise stronger in XX than YY.}
\label{fig:snr}
\end{figure}

\begin{figure}[htbp]
\centering
\subfigure[]{
\includegraphics[width=0.8\linewidth]{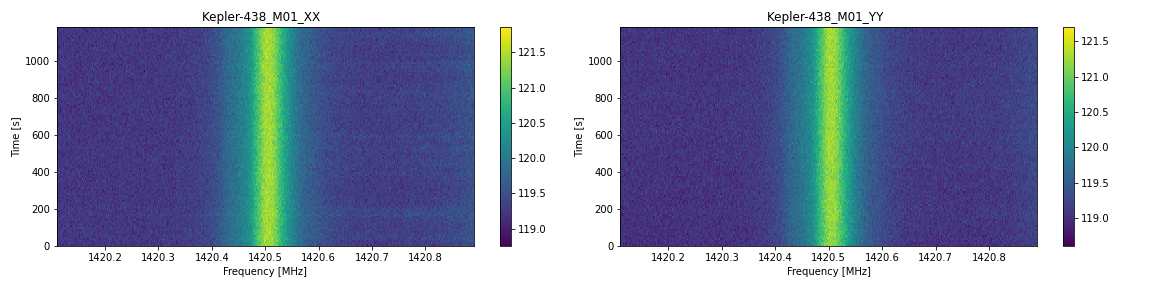}
}
\quad
\subfigure[]{
\includegraphics[width=0.8\linewidth]{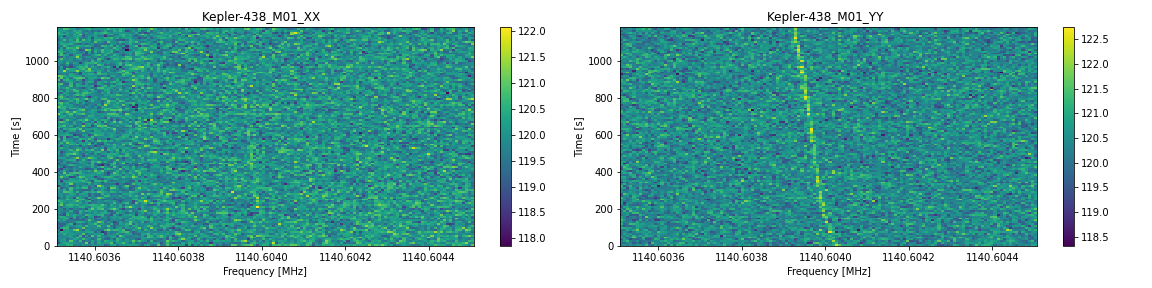}
}
\caption{Waterfall plots of the neutral hydrogen line in the observation towards Kepler-438 (top panels) and NBS 210629 (bottom panels). The left panels show XX and the right panels show YY. The strength of XX and YY of neutral hydrogen line is similar, while NBS 210629 appears obviously stronger in YY than XX.}
\label{fig:HI}
\end{figure}

However, narrow-band signals with unbalanced intensities in XX and YY data are not necessarily sky localized ETI signals, but are also very likely to be instrumental RFI. Because the data of X and Y directions are sampled by the ADC separately in the Roach 2 enclosure, instrumental RFI is likely to affect the data of X and Y directions differently, resulting in different intensities in XX and YY data. To verify this phenomenon, we measure the XX and YY S/Ns of 298 clock oscillator harmonics we find in Beam 1, including a lot of weak signals that are not detected by the program but are visible on the waterfall plots. As shown in Figure \ref{fig:snr}, the majority of these signals display ``polarized'' characteristics: the intensity in one polarization direction is stronger than the other. In addition, the bias depends on frequency, which implies that instrumental RFI signals from the same source have similar polarization characteristics.


To identify the event detected at 1140.604 MHz from the observation towards Kepler-438 (hereafter NBS 210629, where ``NBS'' means narrow-band signal and ``210629'' is the date we detected it), we calculate its S/N more accurately by de-drifting, yielding 9.60 in XX and 22.64 in YY. Compared with the neutral hydrogen line in the same observation, NBS 210629 is apparently stronger in YY than XX (Figure \ref{fig:HI}(b)). Since the 20-minute observation time is relatively short, we cannot find significant variation in the relative intensity between XX and YY. 

\begin{figure}[htbp]
\centering
\includegraphics[width=0.8\linewidth]{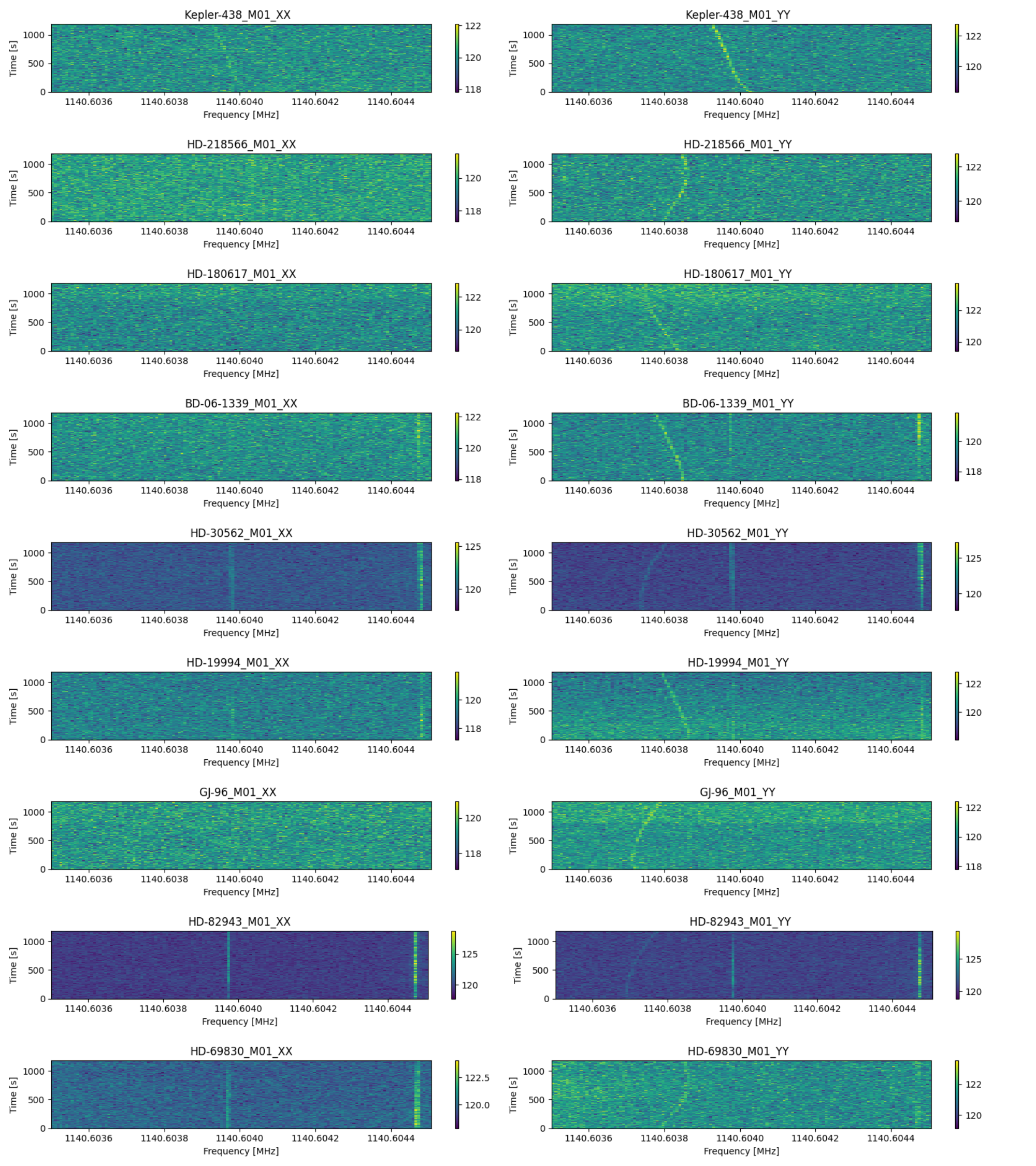}
\caption{Waterfall plots around 1140.604 MHz of the observations towards Kepler-438 and other 8 sources. The left panels show XX and the right panels show YY. The 8 weak signals we find for comparison with NBS 210629 are the drifting ones visible on the right panels. They are only present in Beam 1 and YY, consistent with NBS 210629.}
\label{fig:eight}
\end{figure}

Among the observations towards other targets, we examine the waterfall plots at $1140.604~\mathrm{MHz} \pm 2 ~\mathrm{kHz}$ visually, and 8 narrow-band drifting signals are found (Figure \ref{fig:eight}), which are too weak to reach the S/N threshold we set. Though the frequencies of the 8 signals are slightly lower than that of NBS 210629, and their drift behaviors are different, we find that all the 8 signals are much stronger in YY than XX, and only visible in Beam 1, which is highly consistent with the characteristics of NBS 210629. Therefore, NBS 210629 is most likely to come from the same instrumental source with these 8 signals, but their exact origin is still unknown.

\section{Discussion} \label{sec: Discussion}
\subsection{Sensitivity}

The sensitivity of a radio SETI observation is determined primarily by the effective collecting area and system noise of the telescope. The performance of a telescope can be measured by the system equivalent flux density (SEFD):
\begin{equation}
    \text{SEFD} = \frac{2k_{\text{B}}T_{\text{sys}}}{A_{\text{eff}}},
\end{equation}
where $k_{\text{B}}$ is the Boltzmann constant, $T_{\text{sys}}$ is the system temperature and $A_{\text{eff}}$ is the effective collecting area. The ratio $A_{\text{eff}}/T_{\text{sys}}$ is also known as the sensitivity of the telescope. For the FAST L-band 19-beam receiver, the sensitivity is reported to be $\sim 2000$ m$^{2}$/K \citep{2011IJMPD..20..989N,2016RaSc...51.1060L,2019SCPMA..6259502J}. 

For observations aimed at extremely narrow-band signals, the signal bandwidth is narrower than the frequency resolution. In this case, the minimum detectable flux $S_{\text{min}}$ is given by
\begin{equation}
    S_{\text{min}} = \sigma_{\text{min}}~ \text{SEFD} \sqrt{\frac{\delta\nu}{n_{\text{pol}} ~ t_{\text{obs}}}},
\end{equation}
where $\sigma_{\text{min}}$ is the S/N threshold, $\delta\nu$ is the frequency resolution, $n_{\text{pol}}$ is the {polarization number, and $t_{\text{obs}}$ is the observation duration \citep{2017ApJ...849..104E}. Since we search the data of XX and YY separately, the polarization number is 1. It is worth noting that this expression is different from the $S_{\text{min}}$ for observations aimed at astrophysical signals, where the signal bandwidth is wider than the frequency resolution, so the unit of $S_{\text{min}}$ here is W/m$^{2}$ rather than Jy ($10^{-26}$ W m$^{-2}$ Hz$^{-1}$). Given the S/N threshold of 10, the frequency resolution of $\sim 7.5$ Hz, and the observation duration of 1200 s in this work, we calculate $S_{\text{min}} = 1.09 \times 10^{-26}$ W/m$^{2}$ (1.09 Jy Hz).

Assuming an isotropic transmitter, the distances to the sources should be taken into account when we discuss the sensitivity of targeted radio SETI observations. The minimum detectable equivalent isotropic radiated Power (EIRP$_{\text{min}}$) is defined as
\begin{equation}
    \text{EIRP}_{\text{min}}=4\pi d^{2} S_{\text{min}},
\end{equation}
where $d$ is the distance to the target. The Gaia Early Data Release 3 (EDR3) \citep{2021A&A...649A...1G} shows the closest target in our observations, Ross 128, is 3.37 pc away. Therefore, we calculate our minimum $\text{EIRP}_{\text{min}} = 1.48 \times 10^{9}$ W, which achieves an unprecedented sensitivity. Comparing to the EIRP of the Arecibo Planetary Radar of $\sim 10^{13}$ W, the weakest ETI signal we are able to detect is well within the reach of current human technology.

\subsection{Advantages of MBCM}

The primary goal of the MBCM strategy is to improve the time efficiency of targeted SETI observations. Compared with the on-off strategy, MBCM also shows many significant advantages in addition to its high time efficiency. 

First, MBCM observes more reference locations for RFI identification than the on-off strategy. Studies using the on-off strategy in recent years commonly observe no more than 3 off-sources for one on-source \citep{2013ApJ...767...94S,2017ApJ...849..104E,2019AJ....157..122P,2020AJ....159...86P,2020AJ....160...29S,2021AJ....161..286T}. In contrast, there are as many as six reference beams used in our observations. Due to the diversity of RFI intensity and incident direction, sometimes an RFI signal detected in on-observations can escape the detection by off-observations accidentally. Observing more reference locations can help reduce the possibility of this contingency.

Second, instrumental RFI can be identified by investigating the beam coverage. As mentioned in Results, the clock oscillator harmonics appear in Beam 1 and Beam 2 simultaneously because the two beams share the same FPGA board. The beam coverage pattern of instrumental RFI is subject to the hardware structure of the backend, thus signals with some specific beam coverage patterns are most likely to be instrumental RFI.

Third but most importantly, the on-off strategy cannot identify RFI with a duty cycle matching the on-off cadence \citep{2021NatAs.tmp..206S}, but MBCM is not bothered by this problem. If an RFI signal happens to emerge when observing the on-source and vanish when observing the off-source, the on-off strategy will fail to reject it. Since MBCM observes the target and the reference locations simultaneously, such coincidence can be avoided, and the possibilities of various vehicles (cars, airplanes, etc.) and low-Earth orbit satellites can be ruled out directly, which are common RFI sources in radio SETI.

MBCM can be applied to any single-dish multi-beam radio telescope as long as the reference beams are far enough from the target beam to ensure that signals from the source cannot be detected by the reference beams. On the basis of MBCM, targeted SETI will be further promoted when simultaneous observations by two or more telescopes at different locations become practicable, which can avoid the influence of deceptive RFI both around the observatory site and from the backend instruments.

\subsection{Complements to Technosignature Verification Methodology}

To verify SETI signals of interest, we propose a novel and convenient approach, investigating the polarization characteristics, which provides us with a confident judgement of NBS 210629. Previous targeted SETI studies have accumulated rich experience in identifying candidate signals and established a systematic methodology. Aiming to compare our approach to the former ones and complement the signal identification methodology, we performed a comprehensive examination on NBS 210629 following the technosignature verification framework summarized by \citet{2021NatAs.tmp..206S}.

\begin{enumerate}
\item We verify that the operation of FAST was functioning correctly on the observation day, and exclude the possibility of packet loss and that the ADC frequency synthesizer lost lock, which are the causes of abnormal drifting. 
\item Via visual inspection of the waterfall plots and integrating the de-drifted spectra, we can confirm that NBS 210629 is totally absent from the reference beams. 
\item Referring to the RFI environment test at the FAST site, NBS 210629 is not within the catalogued RFI bands: civil aviations and navigation satellites \citep{2021RAA....21...18W}. 
\item A typical type of human-made signals with electronic drifts is clock oscillator harmonics. As discussed in Results, rejecting this kind of RFI is not complicated, but the frequency of NBS 210629 is not within the known ranges of the harmonics. As shown in Figure \ref{fig:candidate}, the drift rate of NBS 210629 varies slightly at the beginning, which indicates electronic drift. However, modern transmitters can be programmed to produce any drift rate, so can ETI transmitters, thus this cannot be the primary reason to reject the event. Human-made signals with accelerational drifts are usually transmitted from various vehicles and spacecrafts. Signals from cars, airplanes and low-Earth orbit satellites can be excluded directly by MBCM because they will be detected by the reference beams. However, medium-Earth orbit satellites, geosynchronous satellites and deep-space probes are still potential RFI sources. The trajectories of satellites passing overhead during the observation can be queried by the FAST satellite RFI database \citep{2021RAA....21...18W}, where we find no satellite within 3 degrees from the main lobe of Beam 1 as we observed Kepler-438. The coordinates of deep-space probes can be queried from NASA Horizons \footnote{\url{https://ssd.jpl.nasa.gov/horizons.cgi}}, where we find no probe that overlapped with Kepler-438. However, we cannot have access to the information of military satellites, so this procedure cannot absolutely rule out the possibility of human-made spacecrafts.
\item Assuming a ground-based transmitter, the relative acceleration causing the Doppler drift is mainly composed of the rotation and orbit of the Earth and the planet. The maximum drift rate at $\nu_{0}=1.45$ GHz caused by the Earth's rotation at the latitude of FAST is
\begin{equation}
    \dot{\nu}_{\oplus,\text{r}} = \frac{\nu_{0}}{c} \frac{4\pi^{2}R_{\oplus}}{P_{\oplus}^{2}} \cos{\phi} = 0.15 ~\text{Hz/s}.
\end{equation}
The maximum drift rate at $\nu_{0}=1.45$ GHz caused by the Earth's orbit is
\begin{equation}
    \dot{\nu}_{\oplus,\text{o}} = \frac{\nu_{0}}{c} \frac{GM_{\odot}}{r_{\oplus}^{2}} = 0.03 ~\text{Hz/s}.
\end{equation}
The maximum drift rate caused by the planet's orbit is
\begin{equation}
    \dot{\nu}_{\text{p},\text{o}} = \frac{\nu_{0}}{c} \frac{GM_{\star}}{r_{\text{p}}^{2}}.
\end{equation}
For Kepler-438 b, given the orbit semi-major axis of 0.166 AU and the stellar mass of 0.544 $M_{\odot}$ \citep{2015ApJ...800...99T}, the maximum drift rate at $\nu_{0}=1.45$ GHz caused by the orbit is 0.57 Hz/s. So far we are not able to obtain the rotational period of Kepler-438 b, but a planet in the habitable zone around an M-type star is most likely to be tidally locked. These calculations are sufficient to verify that the drift rate of NBS 210629, $-0.0678$ Hz/s, is celestial-mechanically attainable. The calculation of an exact drift rate is complicated, which requires detailed information of the orbital elements and rotational parameters. Actually, such calculation is not necessary, because even if the drift rate of a candidate signal is out of the range derived from celestial mechanics, this will not invalidate the signal because we do not know the exact motion of the transmitter (ground-based or space-based) and whether the signal drifts electronically. In other words, drift rate analyses can give support to a candidate signal, but cannot be the only evidence to reject a candidate signal.
\item Corresponding to the steps 6-9 in \citet{2021NatAs.tmp..206S}, we search our data for RFI signals with similar characteristics to NBS 210629, which may have the same origin with the event. Because a signal may be transmitted at several different frequencies simultaneously, and the frequency of a signal may always be constant, our search is carried out in two directions: (1) RFI with the same frequency-normalized drift rate $\dot{\nu}/\nu$ in the same observation; (2) RFI at the same frequency in other observations. For the former case, though we find some RFI signals with $\dot{\nu}/\nu$ close to NBS 210629, none of them has the same morphology as NBS 210629, which means that their drifts are not synchronous, thus they are unlikely to have the same source. For the latter case, we find the 8 weak signals mentioned in Results within the range of 1140.604 MHz $\pm$ 2 kHz. All the 8 signals are only visible in Beam 1 and YY is significantly stronger than XX, which is highly consistent with the polarization characteristics of NBS 210629, so we can infer that they are most likely to originate from the same source.
\item Using the MBCM strategy, we conducted re-observations towards Kepler-438 on November 7, 2021, March 10, 19 and 28 and April 6, 2022. We did not detect any narrow-band drifting signal at 1140.604 MHz $\pm$ 2 kHz.
\end{enumerate}

Based on our practical experience, we propose to complement this signal verification framework to apply it to MBCM. If a narrow-band technosignature signal of interest is detected, after checking for catalogued RFI at the same frequency at the observatory, we suggest searching for the signal at the same frequency and investigating the polarization characteristics of the signal first before searching for human-made sources emitting signals with similar accelerational or electronic drifts, because this search is very laborious and cannot be performed thoroughly due to too many unknown factors. Polarization characteristics are strong evidences for rejecting instrumental RFI, and the following sophisticated procedures can be omitted if such investigation works. This approach can greatly improve the efficiency of signal verification, and can effectively recognize deceptive instrumental RFI.

\section{Conclusions} \label{sec:con}

We conduct searches for narrow-band drifting radio signals towards 33 exoplanet systems, using the largest single-dish radio telescope in the world. The observed targets include stars hosting planets in their habitable zones and stars in the Earth transit zone, which are targets of great SETI value.

To conduct the observations efficiently and distinguish candidate ETI signals from human-made RFI, we design an observation strategy called multi-beam coincidence matching for FAST. Recording the data on the spectral line backend with a frequency resolution of $\sim$7.5 Hz, we search narrow-band signals across $1.05-1.45$ GHz with drift rates within $\pm 4$ Hz/s and S/Ns above 10, in two orthogonal linear polarization directions separately. 

The vast majority of the detected events are obvious RFI, either false positives or clock oscillator harmonics. The most particular signal, NBS 210629, is identified by its polarization characteristics, which play a key role to reveal that the event is very unlikely to have an extraterrestrial origin. In conclusion, our observations find no solid evidence for 100\%-duty cycle radio transmitters emitting between 1.05 and 1.45 GHz with an EIRP above $1.48 \times 10^{9} ~\text{W}$.

To achieve all the key science goals, FAST will conduct more SETI campaigns in the future, including both targeted searches and commensal sky surveys. Using the most sensitive single-dish radio telescope, SETI with FAST is also challenged by more types of weak RFI. As the SETI campaigns of FAST progress, we will identify and classify more varieties of RFI, especially instrumental RFI with specific beam coverage patterns, and develop new methods to remove them. Meanwhile, we will improve the instrumental RFI shielding ability of the backend to enhance the SETI performance of FAST.


\vspace*{2mm} 

We sincerely appreciate the referee’s suggestions, which help us greatly improve our manuscript. This work was supported by the Ministry of Science and Technology of China, the National Science Foundation of China (NSFC) Programs Grants No. 11929301 and National Key R\&D Program of China (2017YFA0402600), the National Natural Science Foundation of China Grants No. 12041304, CAS ``Light of West China'' Program, Cultivation Project for FAST Scientific Payoff and Research Achievement of CAMS-CAS. This work made use of data from FAST, a Chinese national mega-science facility built and operated by the National Astronomical Observatories, Chinese Academy of Sciences. We sincerely thank Pei Wang for his kind and useful discussions.
\bibliography{text}

\begin{thebibliography}{}
\expandafter\ifx\csname natexlab\endcsname\relax\def\natexlab#1{#1}\fi
\providecommand{\url}[1]{\href{#1}{#1}}
\providecommand{\dodoi}[1]{doi:~\href{http://doi.org/#1}{\nolinkurl{#1}}}
\providecommand{\doeprint}[1]{\href{http://ascl.net/#1}{\nolinkurl{http://ascl.net/#1}}}
\providecommand{\doarXiv}[1]{\href{https://arxiv.org/abs/#1}{\nolinkurl{https://arxiv.org/abs/#1}}}

\bibitem[{{Batalha}(2014)}]{2014PNAS..11112647B}
{Batalha}, N.~M. 2014, Proceedings of the National Academy of Science, 111,
  12647, \dodoi{10.1073/pnas.1304196111}

\bibitem[{{Cocconi} \& {Morrison}(1959)}]{1959Natur.184..844C}
{Cocconi}, G., \& {Morrison}, P. 1959, \nat, 184, 844, \dodoi{10.1038/184844a0}

\bibitem[{{Drake}(1961)}]{1961PhT....14d..40D}
{Drake}, F.~D. 1961, Physics Today, 14, 40, \dodoi{10.1063/1.3057500}

\bibitem[{{Dressing} \& {Charbonneau}(2013)}]{2013ApJ...767...95D}
{Dressing}, C.~D., \& {Charbonneau}, D. 2013, \apj, 767, 95,
  \dodoi{10.1088/0004-637X/767/1/95}

\bibitem[{{Enriquez} \& {Price}(2019)}]{2019ascl.soft06006E}
{Enriquez}, E., \& {Price}, D. 2019, {turboSETI: Python-based SETI search
  algorithm}.
\newblock \doeprint{1906.006}

\bibitem[{{Enriquez} {et~al.}(2017){Enriquez}, {Siemion}, {Foster}, {Gajjar},
  {Hellbourg}, {Hickish}, {Isaacson}, {Price}, {Croft}, {DeBoer}, {Lebofsky},
  {MacMahon}, \& {Werthimer}}]{2017ApJ...849..104E}
{Enriquez}, J.~E., {Siemion}, A., {Foster}, G., {et~al.} 2017, \apj, 849, 104,
  \dodoi{10.3847/1538-4357/aa8d1b}

\bibitem[{{Gaia Collaboration} {et~al.}(2021){Gaia Collaboration}, {Brown},
  {Vallenari}, {Prusti}, {de Bruijne}, {Babusiaux}, {Biermann}, {Creevey},
  {Evans}, {Eyer}, \& et~al.}]{2021A&A...649A...1G}
{Gaia Collaboration}, {Brown}, A.~G.~A., {Vallenari}, A., {et~al.} 2021, \aap,
  649, A1, \dodoi{10.1051/0004-6361/202039657}

\bibitem[{{Gajjar} {et~al.}(2021){Gajjar}, {Perez}, {Siemion}, {Foster},
  {Brzycki}, {Chatterjee}, {Chen}, {Cordes}, {Croft}, {Czech}, {DeBoer},
  {DeMarines}, {Drew}, {Gowanlock}, {Isaacson}, {Lacki}, {Lebofsky},
  {MacMahon}, {Morrison}, {Ng}, {de Pater}, {Price}, {Sheikh}, {Suresh},
  {Webb}, \& {Pete Worden}}]{2021AJ....162...33G}
{Gajjar}, V., {Perez}, K.~I., {Siemion}, A. P.~V., {et~al.} 2021, \aj, 162, 33,
  \dodoi{10.3847/1538-3881/abfd36}

\bibitem[{{Gray} \& {Mooley}(2017)}]{2017AJ....153..110G}
{Gray}, R.~H., \& {Mooley}, K. 2017, \aj, 153, 110,
  \dodoi{10.3847/1538-3881/153/3/110}

\bibitem[{{Han} {et~al.}(2021){Han}, {Wang}, {Wang}, {Wang}, {Zhou}, {Sun},
  {Yan}, {Su}, {Jing}, {Chen}, {Gao}, {Hou}, {Xu}, {Lee}, {Wang}, {Jiang},
  {Xu}, {Yan}, {Gan}, {Guan}, {Huang}, {Jiang}, {Li}, {Men}, {Sun}, {Wang},
  {Wang}, {Wang}, {Xie}, {Xu}, {Yao}, {You}, {Yu}, {Yuan}, {Yuen}, {Zhang}, \&
  {Zhu}}]{2021RAA....21..107H}
{Han}, J.~L., {Wang}, C., {Wang}, P.~F., {et~al.} 2021, Research in Astronomy
  and Astrophysics, 21, 107, \dodoi{10.1088/1674-4527/21/5/107}

\bibitem[{{Harp} {et~al.}(2016){Harp}, {Richards}, {Tarter}, {Dreher},
  {Jordan}, {Shostak}, {Smolek}, {Kilsdonk}, {Wilcox}, {Wimberly}, {Ross},
  {Barott}, {Ackermann}, \& {Blair}}]{2016AJ....152..181H}
{Harp}, G.~R., {Richards}, J., {Tarter}, J.~C., {et~al.} 2016, \aj, 152, 181,
  \dodoi{10.3847/0004-6256/152/6/181}

\bibitem[{{Horowitz} \& {Sagan}(1993)}]{1993ApJ...415..218H}
{Horowitz}, P., \& {Sagan}, C. 1993, \apj, 415, 218, \dodoi{10.1086/173157}

\bibitem[{{Jiang} {et~al.}(2019){Jiang}, {Yue}, {Gan}, {Yao}, {Li}, {Pan},
  {Sun}, {Yu}, {Liu}, {Tang}, {Qian}, {Lu}, {Yan}, {Peng}, {Zhang}, {Wang},
  {Li}, \& {Li}}]{2019SCPMA..6259502J}
{Jiang}, P., {Yue}, Y., {Gan}, H., {et~al.} 2019, Science China Physics,
  Mechanics, and Astronomy, 62, 959502, \dodoi{10.1007/s11433-018-9376-1}

\bibitem[{{Jiang} {et~al.}(2020){Jiang}, {Tang}, {Hou}, {Liu}, {Kr{\v{c}}o},
  {Qian}, {Sun}, {Ching}, {Liu}, {Duan}, {Yue}, {Gan}, {Yao}, {Li}, {Pan},
  {Yu}, {Liu}, {Li}, {Peng}, {Yan}, \& {FAST
  Collaboration}}]{2020RAA....20...64J}
{Jiang}, P., {Tang}, N.-Y., {Hou}, L.-G., {et~al.} 2020, Research in Astronomy
  and Astrophysics, 20, 064, \dodoi{10.1088/1674-4527/20/5/64}

\bibitem[{{Kaltenegger} \& {Faherty}(2021)}]{2021Natur.594..505K}
{Kaltenegger}, L., \& {Faherty}, J.~K. 2021, \nat, 594, 505,
  \dodoi{10.1038/s41586-021-03596-y}

\bibitem[{{Kasting} {et~al.}(1993){Kasting}, {Whitmire}, \&
  {Reynolds}}]{1993Icar..101..108K}
{Kasting}, J.~F., {Whitmire}, D.~P., \& {Reynolds}, R.~T. 1993, \icarus, 101,
  108, \dodoi{10.1006/icar.1993.1010}

\bibitem[{{Keane} {et~al.}(2018){Keane}, {Barr}, {Jameson}, {Morello}, {Caleb},
  {Bhandari}, {Petroff}, {Possenti}, {Burgay}, {Tiburzi}, {Bailes}, {Bhat},
  {Burke-Spolaor}, {Eatough}, {Flynn}, {Jankowski}, {Johnston}, {Kramer},
  {Levin}, {Ng}, {van Straten}, \& {Krishnan}}]{2018MNRAS.473..116K}
{Keane}, E.~F., {Barr}, E.~D., {Jameson}, A., {et~al.} 2018, \mnras, 473, 116,
  \dodoi{10.1093/mnras/stx2126}

\bibitem[{{Kopparapu} {et~al.}(2013){Kopparapu}, {Ramirez}, {Kasting}, {Eymet},
  {Robinson}, {Mahadevan}, {Terrien}, {Domagal-Goldman}, {Meadows}, \&
  {Deshpande}}]{2013ApJ...765..131K}
{Kopparapu}, R.~K., {Ramirez}, R., {Kasting}, J.~F., {et~al.} 2013, \apj, 765,
  131, \dodoi{10.1088/0004-637X/765/2/131}

\bibitem[{{Li} \& {Pan}(2016)}]{2016RaSc...51.1060L}
{Li}, D., \& {Pan}, Z. 2016, Radio Science, 51, 1060,
  \dodoi{10.1002/2015RS005877}

\bibitem[{{Li} {et~al.}(2018){Li}, {Wang}, {Qian}, {Krco}, {Jiang}, {Yue},
  {Jin}, {Zhu}, {Pan}, {Nan}, \& {Dunning}}]{2018IMMag..19..112L}
{Li}, D., {Wang}, P., {Qian}, L., {et~al.} 2018, IEEE Microwave Magazine, 19,
  112, \dodoi{10.1109/MMM.2018.2802178}

\bibitem[{{Li} {et~al.}(2020){Li}, {Gajjar}, {Wang}, {Siemion}, {Zhang},
  {Zhang}, {Yue}, {Zhu}, {Jin}, {Li}, {Berger}, {Brzycki}, {Cobb}, {Croft},
  {Czech}, {DeBoer}, {DeMarines}, {Drew}, {Emilio Enriquez}, {Gizani},
  {Korpela}, {Isaacson}, {Lebofsky}, {Lacki}, {MacMahon}, {Nanez}, {Niu},
  {Pei}, {Price}, {Werthimer}, {Worden}, {Gerry Zhang}, {Zhang}, \& {FAST
  Collaboration}}]{2020RAA....20...78L}
{Li}, D., {Gajjar}, V., {Wang}, P., {et~al.} 2020, Research in Astronomy and
  Astrophysics, 20, 078, \dodoi{10.1088/1674-4527/20/5/78}

\bibitem[{{Li} {et~al.}(2021){Li}, {Wang}, {Zhu}, {Zhang}, {Zhang}, {Duan},
  {Zhang}, {Feng}, {Tang}, {Chatterjee}, {Cordes}, {Cruces}, {Dai}, {Gajjar},
  {Hobbs}, {Jin}, {Kramer}, {Lorimer}, {Miao}, {Niu}, {Niu}, {Pan}, {Qian},
  {Spitler}, {Werthimer}, {Zhang}, {Wang}, {Xie}, {Yue}, {Zhang}, {Zhi}, \&
  {Zhu}}]{2021Natur.598..267L}
{Li}, D., {Wang}, P., {Zhu}, W.~W., {et~al.} 2021, \nat, 598, 267,
  \dodoi{10.1038/s41586-021-03878-5}

\bibitem[{{Lin} {et~al.}(2020){Lin}, {Zhang}, {Wang}, {Gao}, {Guan}, {Han},
  {Jiang}, {Jiang}, {Lee}, {Li}, {Men}, {Miao}, {Niu}, {Niu}, {Sun}, {Wang},
  {Wang}, {Xu}, {Xu}, {Xu}, {Yang}, {Yang}, {Yu}, {Zhang}, {Zhang}, {Zhou},
  {Zhu}, {Castro-Tirado}, {Dai}, {Ge}, {Hu}, {Li}, {Li}, {Li}, {Liang}, {Jia},
  {Querel}, {Shao}, {Wang}, {Wang}, {Wu}, {Xiong}, {Xu}, {Yang}, {Zhang},
  {Zhang}, {Zheng}, \& {Zou}}]{2020Natur.587...63L}
{Lin}, L., {Zhang}, C.~F., {Wang}, P., {et~al.} 2020, \nat, 587, 63,
  \dodoi{10.1038/s41586-020-2839-y}

\bibitem[{{MacMahon} {et~al.}(2018){MacMahon}, {Price}, {Lebofsky}, {Siemion},
  {Croft}, {DeBoer}, {Enriquez}, {Gajjar}, {Hellbourg}, {Isaacson},
  {Werthimer}, {Abdurashidova}, {Bloss}, {Brandt}, {Creager}, {Ford}, {Lynch},
  {Maddalena}, {McCullough}, {Ray}, {Whitehead}, \&
  {Woody}}]{2018PASP..130d4502M}
{MacMahon}, D. H.~E., {Price}, D.~C., {Lebofsky}, M., {et~al.} 2018, \pasp,
  130, 044502, \dodoi{10.1088/1538-3873/aa80d2}

\bibitem[{{Nan}(2006)}]{2006ScChG..49..129N}
{Nan}, R. 2006, Science in China: Physics, Mechanics and Astronomy, 49, 129,
  \dodoi{10.1007/s11433-006-0129-9}

\bibitem[{{Nan} {et~al.}(2011){Nan}, {Li}, {Jin}, {Wang}, {Zhu}, {Zhu},
  {Zhang}, {Yue}, \& {Qian}}]{2011IJMPD..20..989N}
{Nan}, R., {Li}, D., {Jin}, C., {et~al.} 2011, International Journal of Modern
  Physics D, 20, 989, \dodoi{10.1142/S0218271811019335}

\bibitem[{{NASA Exoplanet Archive}(2019)}]{https://doi.org/10.26133/nea1}
{NASA Exoplanet Archive}. 2019, Confirmed Planets Table,  IPAC,
  \dodoi{10.26133/NEA1}

\bibitem[{{PEI} {et~al.}(2019){PEI}, {LI}, {LI}, \&
  {NIU}}]{2019SSPMA..49i9508P}
{PEI}, X., {LI}, J., {LI}, S., \& {NIU}, C. 2019, Scientia Sinica Physica,
  Mechanica \& Astronomica, 49, 099508, \dodoi{10.1360/SSPMA2018-00418}

\bibitem[{{Petigura} {et~al.}(2013){Petigura}, {Howard}, \&
  {Marcy}}]{2013PNAS..11019273P}
{Petigura}, E.~A., {Howard}, A.~W., \& {Marcy}, G.~W. 2013, Proceedings of the
  National Academy of Science, 110, 19273, \dodoi{10.1073/pnas.1319909110}

\bibitem[{{Pinchuk} {et~al.}(2019){Pinchuk}, {Margot}, {Greenberg}, {Ayalde},
  {Bloxham}, {Boddu}, {Gerardo Chinchilla-Garcia}, {Cliffe}, {Gallagher},
  {Hart}, {Hesford}, {Mizrahi}, {Pike}, {Rodger}, {Sayki}, {Schneck}, {Tan},
  {{\textquotedblleft}Yolanda{\textquotedblright} Xiao}, \&
  {Lynch}}]{2019AJ....157..122P}
{Pinchuk}, P., {Margot}, J.-L., {Greenberg}, A.~H., {et~al.} 2019, \aj, 157,
  122, \dodoi{10.3847/1538-3881/ab0105}

\bibitem[{{Price} {et~al.}(2019){Price}, {Enriquez}, {Chen}, \&
  {Siebert}}]{2019JOSS....4.1554P}
{Price}, D., {Enriquez}, J., {Chen}, Y., \& {Siebert}, M. 2019, The Journal of
  Open Source Software, 4, 1554, \dodoi{10.21105/joss.01554}

\bibitem[{{Price} {et~al.}(2020){Price}, {Enriquez}, {Brzycki}, {Croft},
  {Czech}, {DeBoer}, {DeMarines}, {Foster}, {Gajjar}, {Gizani}, {Hellbourg},
  {Isaacson}, {Lacki}, {Lebofsky}, {MacMahon}, {Pater}, {Siemion}, {Werthimer},
  {Green}, {Kaczmarek}, {Maddalena}, {Mader}, {Drew}, \&
  {Worden}}]{2020AJ....159...86P}
{Price}, D.~C., {Enriquez}, J.~E., {Brzycki}, B., {et~al.} 2020, \aj, 159, 86,
  \dodoi{10.3847/1538-3881/ab65f1}

\bibitem[{{Qian} {et~al.}(2019){Qian}, {Pan}, {Li}, {Hobbs}, {Zhu}, {Wang},
  {Liu}, {Yue}, {Zhu}, {Liu}, {Yu}, {Sun}, {Jiang}, {Pan}, {Li}, {Gan}, {Yao},
  {Xie}, {Camilo}, {Cameron}, {Zhang}, \& {Wang}}]{2019SCPMA..6259508Q}
{Qian}, L., {Pan}, Z., {Li}, D., {et~al.} 2019, Science China Physics,
  Mechanics, and Astronomy, 62, 959508, \dodoi{10.1007/s11433-018-9354-y}

\bibitem[{{Sheikh} {et~al.}(2020){Sheikh}, {Siemion}, {Enriquez}, {Price},
  {Isaacson}, {Lebofsky}, {Gajjar}, \& {Kalas}}]{2020AJ....160...29S}
{Sheikh}, S.~Z., {Siemion}, A., {Enriquez}, J.~E., {et~al.} 2020, \aj, 160, 29,
  \dodoi{10.3847/1538-3881/ab9361}

\bibitem[{{Sheikh} {et~al.}(2021){Sheikh}, {Smith}, {Price}, {DeBoer}, {Lacki},
  {Czech}, {Croft}, {Gajjar}, {Isaacson}, {Lebofsky}, {MacMahon}, {Ng},
  {Perez}, {Siemion}, {Webb}, {Zic}, {Drew}, \& {Worden}}]{2021NatAs.tmp..206S}
{Sheikh}, S.~Z., {Smith}, S., {Price}, D.~C., {et~al.} 2021, Nature Astronomy,
  \dodoi{10.1038/s41550-021-01508-8}

\bibitem[{{Siemion} {et~al.}(2013){Siemion}, {Demorest}, {Korpela},
  {Maddalena}, {Werthimer}, {Cobb}, {Howard}, {Langston}, {Lebofsky}, {Marcy},
  \& {Tarter}}]{2013ApJ...767...94S}
{Siemion}, A. P.~V., {Demorest}, P., {Korpela}, E., {et~al.} 2013, \apj, 767,
  94, \dodoi{10.1088/0004-637X/767/1/94}

\bibitem[{{Smith} {et~al.}(2021){Smith}, {Price}, {Sheikh}, {Czech}, {Croft},
  {DeBoer}, {Gajjar}, {Isaacson}, {Lacki}, {Lebofsky}, {MacMahon}, {Ng},
  {Perez}, {Siemion}, {Webb}, {Drew}, {Worden}, \& {Zic}}]{2021NatAs.tmp..203S}
{Smith}, S., {Price}, D.~C., {Sheikh}, S.~Z., {et~al.} 2021, Nature Astronomy,
  \dodoi{10.1038/s41550-021-01479-w}

\bibitem[{{Tarter}(2001)}]{2001ARA&A..39..511T}
{Tarter}, J. 2001, \araa, 39, 511, \dodoi{10.1146/annurev.astro.39.1.511}

\bibitem[{{Tarter} {et~al.}(1980){Tarter}, {Cuzzi}, {Black}, \&
  {Clark}}]{1980Icar...42..136T}
{Tarter}, J., {Cuzzi}, J., {Black}, D., \& {Clark}, T. 1980, \icarus, 42, 136,
  \dodoi{10.1016/0019-1035(80)90251-1}

\bibitem[{{Torres} {et~al.}(2015){Torres}, {Kipping}, {Fressin}, {Caldwell},
  {Twicken}, {Ballard}, {Batalha}, {Bryson}, {Ciardi}, {Henze}, {Howell},
  {Isaacson}, {Jenkins}, {Muirhead}, {Newton}, {Petigura}, {Barclay},
  {Borucki}, {Crepp}, {Everett}, {Horch}, {Howard}, {Kolbl}, {Marcy},
  {McCauliff}, \& {Quintana}}]{2015ApJ...800...99T}
{Torres}, G., {Kipping}, D.~M., {Fressin}, F., {et~al.} 2015, \apj, 800, 99,
  \dodoi{10.1088/0004-637X/800/2/99}

\bibitem[{{Traas} {et~al.}(2021){Traas}, {Croft}, {Gajjar}, {Isaacson},
  {Lebofsky}, {MacMahon}, {Perez}, {Price}, {Sheikh}, {Siemion}, {Smith},
  {Drew}, \& {Worden}}]{2021AJ....161..286T}
{Traas}, R., {Croft}, S., {Gajjar}, V., {et~al.} 2021, \aj, 161, 286,
  \dodoi{10.3847/1538-3881/abf649}

\bibitem[{{Valdes} \& {Freitas}(1986)}]{1986Icar...65..152V}
{Valdes}, F., \& {Freitas}, R.~A., J. 1986, \icarus, 65, 152,
  \dodoi{10.1016/0019-1035(86)90069-2}

\bibitem[{{Verschuur}(1973)}]{1973Icar...19..329V}
{Verschuur}, G.~L. 1973, \icarus, 19, 329, \dodoi{10.1016/0019-1035(73)90109-7}

\bibitem[{{Wang} {et~al.}(2021){Wang}, {Zhang}, {Hu}, {Huang}, {Zhu}, {Zhi},
  {Zhang}, {Fan}, \& {Yang}}]{2021RAA....21...18W}
{Wang}, Y., {Zhang}, H.-Y., {Hu}, H., {et~al.} 2021, Research in Astronomy and
  Astrophysics, 21, 018, \dodoi{10.1088/1674-4527/21/1/18}

\bibitem[{{Werthimer} {et~al.}(2001){Werthimer}, {Anderson}, {Bowyer}, {Cobb},
  {Heien}, {Korpela}, {Lampton}, {Lebofsky}, {Marcy}, {McGarry}, \&
  {Treffers}}]{2001SPIE.4273..104W}
{Werthimer}, D., {Anderson}, D., {Bowyer}, C.~S., {et~al.} 2001, in Society of
  Photo-Optical Instrumentation Engineers (SPIE) Conference Series, Vol. 4273,
  The Search for Extraterrestrial Intelligence (SETI) in the Optical Spectrum
  III, ed. S.~A. {Kingsley} \& R.~{Bhathal}, 104--109,
  \dodoi{10.1117/12.435384}

\bibitem[{{Williams} \& {Pollard}(2002)}]{2002IJAsB...1...61W}
{Williams}, D.~M., \& {Pollard}, D. 2002, International Journal of
  Astrobiology, 1, 61, \dodoi{10.1017/S1473550402001064}

\bibitem[{{Zhang}(2020)}]{2020FrPhy..1554502Z}
{Zhang}, B. 2020, Frontiers of Physics, 15, 54502,
  \dodoi{10.1007/s11467-020-0973-5}

\bibitem[{{Zhang} {et~al.}(2020){Zhang}, {Werthimer}, {Zhang}, {Cobb},
  {Korpela}, {Anderson}, {Gajjar}, {Lee}, {Li}, {Pei}, {Zhang}, {Huang},
  {Wang}, {Zhu}, {Duan}, {Zhang}, {Jin}, {Zhu}, \& {Li}}]{2020ApJ...891..174Z}
{Zhang}, Z.-S., {Werthimer}, D., {Zhang}, T.-J., {et~al.} 2020, \apj, 891, 174,
  \dodoi{10.3847/1538-4357/ab7376}

\bibitem[{{Zhu} {et~al.}(2020){Zhu}, {Li}, {Luo}, {Miao}, {Zhang}, {Spitler},
  {Lorimer}, {Kramer}, {Champion}, {Yue}, {Cameron}, {Cruces}, {Duan}, {Feng},
  {Han}, {Hobbs}, {Niu}, {Niu}, {Pan}, {Qian}, {Shi}, {Tang}, {Wang}, {Wang},
  {Yuan}, {Zhang}, {Zhang}, {Cao}, {Feng}, {Gan}, {Gao}, {Gu}, {Guo}, {Hao},
  {Huang}, {Huang}, {Jiang}, {Jin}, {Li}, {Li}, {Li}, {Liu}, {Pan}, {Peng},
  {Qian}, {Shi}, {Song}, {Song}, {Sun}, {Sun}, {Wang}, {Wang}, {Wang}, {Xie},
  {Yan}, {Yang}, {Yang}, {Yao}, {Yu}, {Yu}, {Zhang}, {Zhang}, {Zhang}, {Zheng},
  {Zhou}, {Zhu}, {Zhu}, {Zhu}, {Zhu}, \& {Zhu}}]{2020ApJ...895L...6Z}
{Zhu}, W., {Li}, D., {Luo}, R., {et~al.} 2020, \apjl, 895, L6,
  \dodoi{10.3847/2041-8213/ab8e46}

\end{thebibliography}



\end{document}